\documentclass[12pt,A14]{JHEP3}

\usepackage{amssymb, amsmath, amsopn, amsthm}

\usepackage{epsfig}

\usepackage{graphics}

\title{On The  Stability Of  Non-Supersymmetric AdS Vacua}

\author{Prithvi Narayan, ~and Sandip P. Trivedi,

~\\

Tata Institute for Fundamental Research \\
Mumbai 400005, India\\

\vspace{0.1cm}

\email{Email: prithvi.narayan@gmail.com,  trivedi.sp@gmail.com}\\

}

\abstract{
  We consider two infinite families of Non-Supersymmetric $AdS_4$ vacua, called Type 2) and Type 3) vacua, 
 that arise in massive IIA supergravity with flux. We show that both families  are  perturbatively stable. 
We then examine  non-perturbative decays of these vacua to other supersymmetric and non-supersymmetric $AdS_4$ vacua
 mediated by  instantons in the thin wall approximation. We find that many  decays are ruled out since the tension of the interpolating domain wall is too big compared to the energy difference in AdS units. 
In fact,   within our approximations no decays of Type 2) vacua are allowed, although some  decays are only marginally forbidden. 
This can be understood in terms of a ``pairing symmetry" in the landscape which relate Type 2) vacua with supersymmetric ones of the same 
energy. 
}

\preprint{ TIFR/TH/10-05}

\def\beq{\begin{equation}}

\def\eeq{\end{equation}}

\def\bea{\begin{eqnarray}}

\def\eea{\end{eqnarray}}

\begin{document}

\tableofcontents

\section{Introduction}

String Theory has a rich and  complicated landscape of vacua. Non-supersymmetric  anti-deSitter ($AdS$) vacua are an interesting class
amongst these. Lacking supersymmetry, they are not as well understood as  supersymmetric spacetimes. 
But being time independent,  they  should be easier to understand than deSitter spacetimes. 

In this paper we construct a class of non-supersymmetric AdS vacua which are perturbatively stable and investigate 
their non-perturbative stability. 

Our construction is based on massive IIA theory 
compactified on a particular Calabi-Yau manifold. After suitably orientifolding and adding fluxes one 
obtains vacua where all moduli are stabilized and supersymmetry
is broken with a negative cosmological constant. The Calabi-Yau manifold is obtained by blowing-up a $T^6/(Z_3\times Z_3)$
orbifold.  This model was studied in considerable depth by \cite{DGKT}. Here we carefully include the effects of the blow-up modes and 
related fluxes and analyze the stability of the resulting non-susy vacua in detail. 

We find  two  classes of non-susy vacua 
which are perturbatively stable. These are called  Type 2) and Type 3) vacua in our terminology (Type 1 vacua are supersymmetric).
 Both classes contain an infinite number of vacua. 

Next we turn to a study of the non-perturbative stability of these vacua. 
In fact, such an investigation  was one of the main motivations for this paper. 
At first sight one might expect that a  non-supersymmetric AdS vacua in the landscape  would  always  have some non-zero  
rate to decay to 
other  vacua with  lower vacuum energy. Such a decay rate, if it is small enough, 
would not have very  drastic consequences for an observer in AdS space. 

However,  from the point of view of a dual CFT the small  decay rate in the bulk leads to a divergence and 
has dramatic repercussions. 
For, consider  $AdS_{d+1}$  in Poincare coordinates:
\begin{equation}
ds^2=({r\over R})^2 (-dt^2 +\sum_{i=1, \cdots, d-1}(dx^i)^2) + ({R\over r})^2 dr^2.   
\end{equation}
Let the decay rate per unit volume in the bulk be $\Gamma$. 
The corresponding decay rate  per unit volume in the boundary is 
obtained by integrating the bulk decay rate in the radial direction. 
Taking the boundary metric to be flat, 
 the decay rate per unit volume in the boundary theory is, 
\begin{equation}
\label{bdecay}
\Gamma_{boundary}=\int \sqrt{g} dr \Gamma \sim r_{boundary}^{d+1} \Gamma, 
\end{equation}
where $r_{boundary}$ is the radial  location of  the boundary. As $r_{boundary}$ is taken to infinity we see that this diverges. 

Thus an arbitrarily small decay rate in the bulk leads to an infinitely fast decay in the boundary. The putative CFT dual meets with 
an instantaneous end  and cannot exist. 
This consequence of a bulk decay was noted in \cite{HOP} where the decay rate in a non-susy AdS spacetime, obtained by taking an orbifold
of $AdS_5\times S^5$, was discussed in some detail \footnote{Another way to state this result is that the CFT has no scale 
in it so the only answer for a decay rate 
could be either zero or infinite. Non-zero $\Gamma$ leads to an infinite result.} 
\footnote{In the discussion leading upto eq.(\ref{bdecay}) above we have kept the boundary volume 
fixed and not scaled it with the cut-off $r_{boundary}$. This is the standard convention in studying 
AdS/CFT,  e.g., in global coordinates the boundary $S^3$ is taken to be of fixed radius. However, for 
purposes of studying the decay in the boundary theory, it might be 
 more transparent to scale the volume with 
$r_{boundary}$. The discussion above then suggests that the boundary theory also has a finite decay rate 
per unit volume, and the divergence in $\Gamma_{boundary}$ is really a consequence of summing over 
the volume in the boundary theory which diverges as $r_{boundary} \rightarrow \infty$.  We thank A. Sen for these comments.}.

We see  then that our expectation that the non-susy AdS vacua  are  unstable non-perturbatively suggests that  
 non-supersymmetric CFT's which admit a gravity dual are unlikely to  exist \footnote{Of course non-supersymmetric CFT's clearly do exist.
The worry here is about CFT's which admit a gravity dual.}. If true,  this is an important consequence 
since  holography has emerged as a major tool with which to study strongly coupled conformal field theories. 

The two large families of perturbatively stable non-susy vacua  mentioned above provide us with 
 a laboratory in which we  can investigate this  issue of non-perturbative stability. 
Somewhat surprisingly, our analysis reveals that in a large number of cases the decays  to vacua with lower energy 
 are in fact  ruled out.  

The essential reason for this is  the geometry of AdS space. 
 The dynamics of a decay is governed by competition between the volume gain in bulk energy and the surface cost due to 
the tension of the interpolating domain wall. 
 For a system in flat space, say water and 
steam\footnote{In this case it is the difference in Free energy rather than energy that is relevant.},
the volume grows more rapidly than the surface area, so eventually  the volume gain always wins out and 
an instanton always exists which allows a transition to the more stable phase. However in AdS space the volume and area grow at the same
 rate. This means in AdS space a decay can only happen if the tension of the interpolating domain wall is small enough compared to the difference
in energies between the two vacua. Explicit calculations show that in several cases  the tension  
turns out to be much too large  thereby forbidding the  decay.

In fact it turns out that  working within our approximations, Type 2) vacua are stable and can at best decay marginally. 
Type 3) vacua, in contrast, do indeed decay to some  other Type 3) and Type 2) vacua.
We find no decays of the non-susy vacua to susy ones are allowed.

It is important to go beyond our approximations especially for decays which are marginal at leading order. 
Subleading corrections should tip the marginal cases one way or another and this would determine the stability of the Type 2) vacua.
We leave such an investigation for the future. 

Among our approximations one of the more significant ones is  
the thin wall approximation, as described in the classic paper 
by Coleman and Deluccia \cite{CDL}. 
The domain walls involved cause a jump in flux and thus carry D-brane charges. At first sight one would expect them to be just D-branes
and therefore  well described
in the thin wall approximation. However  the change in flux also causes the moduli to vary and these typically have a mass of order 
the AdS scale. As a result  the domain walls  are  no longer  thin \footnote{Fat would not be PC.}. 
To work within the thin wall approximation then we must restrict ourselves to cases where the moduli variation contributes little, compared to the 
D-brane contribution, to the tension of the domain wall. This allows the  domain wall to  be approximated as being just a   D-brane, which 
is indeed thin. 
Another limitation comes from not having explored the full set of non-susy vacua \footnote{In particular 
we have only investigated choices of 
flux for which the Calabi-Yau manifold is a slightly blown-up version of the $T^6/(Z_3\times Z_3)$ orbifold. General non-susy vacua, 
where the moduli could be  stabilized far away from the orbifold point, have not been analyzed.}. 
There could be other non-susy vacua which we have not constructed  which are allowed end  points of decays.

Despite these limitations in our analysis, we find it significant  that a large number of  possible decays are  in fact  
 ruled out. This indicates that perhaps stable non-susy AdS vacua and associated CFT's might exist after all.  

We should also mention that  the stability of Type 2) vacua can be understood in terms of a ``pairing symmetry" in the landscape.
By reversing the sign of the four-form flux in these vacua one obtains  susy vacua with the same vacuum energy.
The stability of the Type 2) vacua then follows from the stability of their partner susy vacua.
While we have established the argument which relates the stability of the Type 2) vacua to that of their susy partners 
only within our approximations it could have a greater range of validity.

The basic strategy we employ in studying the non-perturbative stability is as follows. The tension of the 
interpolating domain wall satisfies a  lower bound in terms of the jump in the superpotential caused by the domain wall. 
By comparing this lower bound against an upper bound which must be met for the decay to be allowed in AdS space, several decays can be ruled out. 

Such a  bound on the tension of domain walls  is familiar in supersymmetric settings. It might seem puzzling at first 
that it arises  in our study of non-susy vacua.  
The essential reason  is that  even in the non-supersymmetric case the  
compactification can be well approximated to be a Calabi-Yau space. The ratio of
the size of the internal space  to the $AdS$ radius  goes to zero for large flux in these compactifications making them of non Freund-Rubin type. 
This means that the main effect of the fluxes in these cases is  to  stabilize the moduli while  the shifts of the Kaluza Klein modes is  small.   
The tension of the domain wall, which is fixed
to good approximation by the geometry of the Calabi-Yau space, is then determined by supersymmetric data and can be bounded by the 
jump in the superpotential. We  expect that a similar strategy should be useful in other flux compactifications as well, where the internal 
compactification can continue to be well approximated as a supersymmetric one. 
 It might also be useful in going beyond the thin wall approximation. 
 
Let us end this introduction by discussing some related  literature.
The idea of  reversing the sign of fluxes to obtain non-susy vacua from susy ones, which we have used for Type 2) cases, is an old one in the sugra literature.  See \cite{Duff1} and references therein for a discussion of ``skew whiffing", and also in the context of black holes, \cite{Dabholkar}, 
\cite{Tripathy}. 
For other recent constructions of  non-susy $AdS_4$  vacua see \cite{PS}, \cite{Gaiotto1}.  
A general review for flux compactifications is 
\cite{DK}, which contains several references. 
Early papers on IIA compactifications include \cite{Louis} which developed the 4D framework, and  \cite{Deren}, \cite{KKP}, which discuss
 Moduli stabilization.  Also see, \cite{Romans}, \cite{Lust}, \cite{addone}, \cite{addtwo}, \cite{Kors}. 
A recent  discussion  on non-perturbative decays,  especially  decays of Minkowski  nearly susy vacua  to AdS space and related topics,
 is in \cite{Dine}. Aspects of metastability in AdS space are discussed in \cite{Harlow}. 
 
This paper is organized as follows. In section II we review the Model  discussed in \cite{DGKT} and  construct  the non-susy vacua. 
In \S3 we briefly review the discussion of vacuum decay in \cite{CDL} with particular emphasis on AdS spacetimes. 
In \S4 we turn to non-perturbative decays and analyze  decays mediated by $D4$-branes. 
This is a long section, subsections \S4.1-\S4.3 contain some of the main points. 
More general decays are briefly discussed  in \S5. 
Appendices A-D contain important supplementary  details.

\section{The Model}
We will consider a simple compactification of massive Type IIA 
theory \cite{Romans} on a slightly blown up $T^6/(Z_3\times Z_3)$ orbifold \cite{Dixon}. By turning on  flux  one can get
   non-susy AdS vacua  with all moduli being stabilized. This model was discussed extensively in  \cite{DGKT}. We will follow their notation for the most part and only discuss the essential features of this
compactification, referring the reader to \cite{DGKT} for further details.
 
We will show in this section 
 that for appropriate choices of flux the vacua do not have any tachyons lying below the BF bound and thus are perturbatively
stable. This will set the stage to consider their possible non-perturbative decays in the following sections. 

Let us begin by describing the   $T^6/(Z_3 \times Z_3)$ orbifold. This is an orbifold limit of a Calabi-Yau three-fold. 
Let $z_i=x_i+iy_i, i=1 \cdots 3$ be  three complex coordinates on the $T^6$.  They satisfy the periodicity conditions, 
\begin{equation}
\label{periods}
z_i\simeq z_i+1, \ \ \ z_i\simeq z_i+ \alpha, \alpha=e^{\pi i \over 3}.
\end{equation}
The first $Z_3$, called $T$,   identifies  points related by the transformation,
\begin{equation}
\label{trans}
(z_1,z_2,z_3) \rightarrow (\alpha^2z_1, \alpha^2 z_2,\alpha^2 z_3).
\end{equation}
The resulting orbifold  has $27$ fixed points. The  second $Z_3$ has the  generator, 
\beq
\label{secondz3}
(z_1,z_2,z_3)\rightarrow (\alpha^2 z_1+{1+\alpha \over 3}, \alpha^4 z_2+{1+\alpha \over 3}, z_3+{1+\alpha \over 3}).
\eeq
This leaves $9$  fixed points. 

The resulting compactification has no complex structure moduli. 
The six -torus is a product manifold, $T^6=T^2\times T^2\times T^2$. 
The resulting orbifold has  three Kahler moduli corresponding to the sizes of the three $T^2$'s. Each $T^2$ also 
gives rise to a zero mode for $B_2$ giving rise to three axions; together these give rise to three complex moduli. 
 The dilaton and  an axion which arises from $C_3$,  give rise to one more complex modulus. 
Finally there are nine complex moduli which arise from metric and $B_2$ moduli associated with the $9$ blow up modes. 

We will need to consider a further  $Z_2$  orientifold of this orbifold. This is obtained by modding out by 
$ {\cal O}= \Omega_p (-1)^{F_L} \sigma$, where $\Omega_P$ is world sheet orientation reversal, $F_L$ is left-moving fermion  number, and $\sigma$ 
is reflection,
\beq
\label{reflection}
\sigma: z_i \rightarrow -{\bar z}_i, i =1,2,3. 
\eeq
There is a single $0_6$ plane which fills the non-compact directions and wraps a $3$-cycle which is the locus of fixed points of the 
 $\sigma$ reflection symmetry. The resulting compactification now  has ${\cal N}=1$ supersymmetry.  
The three $T^2$  moduli, the dilaton-axion, and the nine blow up modes, all survive the orientifolding and form 
the bosonic components of chiral superfields. 

We now turn to incorporating the effects of flux. 
To begin, we  discuss the effects of flux in the orbifold limit. Subsequently we will include the 
blow-up modes and related fluxes as well.

\subsection{Fluxes, Superpotential and Potential}
A basis of two-forms on the three $T^2$'s is given by, 
\beq
\label{basistwo}
\omega_i=(\kappa \sqrt{3})^{1/3} i dz^i\wedge d{\bar z}^i, i = 1 \cdots 3.
\eeq
$\kappa$ which appears in the normalization above is defined later in terms of the triple intersection number in eq.(\ref{triint}).
A basis for dual $4$-forms is,
\beq
\label{basisfour}
\tilde{\omega}_i=({3 \over \kappa})^{1/3}(i dz^j\wedge d{\bar z}^j ) (i dz^k \wedge d{\bar z}^k).
\eeq

The holomorphic three-form is , 
\begin{equation}
\label{holothreeform}
\Omega = 3^{1/4} i dz_1\wedge dz_2\wedge dz_3.
\end{equation}
Its real and imaginary parts are $\alpha_0, \beta_0$,  
\beq
\label{reim}
\Omega={1\over \sqrt{2}} (\alpha_0+i \beta_0).
\eeq
Under the $Z_2$ orientifold symmetry $\alpha_0, \beta_0$ are respectively even and odd.  

The three- form $H_3$  has odd intrinsic particy under ${\cal O}$.  Therefore a three-form  flux,
\beq
\label{threeform}
H_3=-p \beta_0,
\eeq
with $p$ constant, is allowed by the equations of motion and Bianchi identities and the  symmetries. 
Similarly the form form flux $F_4$  can be expanded as\footnote{(we are ignoring the possibility of turning on $F_4$ with components along  $4$-cycles dual to the blow up modes for now, these will also be incorporated
subsequently.)} 
\beq
\label{fourform}
  F_4=e_i \tilde{\omega}_i. 
\eeq

After accounting for the presence of Cherns- Simons terms the tadpole condition for the $C_7$ potential is given by, 
\beq
\label{tadpole}
m_0 p = -2 \sqrt{2} \kappa_{10}^2 \mu_6.
\eeq
Here $m_0$ is the Romans parameter, and $\mu_6$ is the tension of  a single $D_6$ brane (an $O_6$ plane has tension $2 \mu_6$).
We note that in our conventions,  $2 \kappa_{10}^2= (2\pi)^7(\alpha')^4$,  $\mu_6 =(2\pi)^{-6}(\alpha')^{-7/2}$. 

The metric takes the form, 
\beq
\label{metric}
ds^2=\sum_{i=1}^3 \gamma_i dz^i d\bar{z}^i .
\eeq
It is convenient to work with the  moduli, 
\beq
\label{kmoduli}
v_i={1\over 2} {1\over (\kappa \sqrt{3})^{1/3}}\gamma_i.
\eeq
 below. These  combine naturally with the axions coming from $B_2$ to
give the bosonic components of chiral superfields as was mentioned above. 
For now we suppress the dependence on the $B_2$ axions and also the dependence on the axion coming from $C_3$. 

The resulting potential in the $4$ dimensional Einstein frame  effective theory   is  then, 
\beq
\label{potential}
V={p^2 \over 4} {e^{2\phi}\over (vol)^2} + {1\over 2}(\sum_{i=1}^3 e_i^2v_i^2) {e^{4\phi}\over (vol^3)}+{m_0^2\over 2}{e^{4\phi}\over vol}-\sqrt{2}
|m_0 p| {e^{3\phi}\over vol^{3/2}}.
\eeq
where $vol$, which is related to the  volume of the compactification\footnote{After the additional $Z_2$ orientifolding, the  volume of the internal space
becomes ${vol\over 2}$. For a more complete discussion of our conventions see subsection \S2.2.2.}, is defined to be 
\beq
\label{defvol}
vol=\int _{T^6/(Z_3)^2}\sqrt{g_6}=\kappa v_1v_2v_3.
\eeq
The four terms on the rhs of the potential above arise from the $|H_3|^2$, $|F_4|^4$, $m_0^2$ and the tension of the $O_6$ planes respectively. 

The important point for the present analysis is that the potential is an even function of the fluxes $e_i$. Thus the minimum value of the potential 
and the location of the minimum in moduli space will only depend on the absolute values of $e_i$ and not on their signs. 
We emphasize this because the conditions for supersymmetry do care about signs. These conditions take the form, 
\beq
\label{condb}
\text{sign}(m_0 e_i) <0, \text{sign}(m_0p)<0. 
\eeq
The second condition is automatically met once the tadpole condition eq.(\ref{tadpole}) is satisfied. 
It  follows from eq.(\ref{condb})  that the $e_i$'s must have the same sign to preserve susy. 

This gives an easy way to construct non-supersymmetric minima. Starting with the supersymmetric case, we can change the sign of some or all of the 
$e_i$'s (while keeping $m_0, p$ fixed).  The tadpole condition eq.(\ref{tadpole}) will continue to be met but the susy conditions will not be.
However since the potential is an even function of the $e_i$ fluxes, the susy minimum in moduli space will continue to be a minimum even for these
non-susy choice of fluxes.

Now that we have understood the basic idea behind the construction of the non-supersymmetric vacua we turn to exploring them in more detail below. 

\subsection{The Superpotential}
To begin, we discuss the  general case of massive IIA on a $CY_3$ orientifold, and then specialize to the compactification at hand. 
Let $\omega_a, a= 1 , \cdots h^{1,1}$ be a basis of $(1,1)$ forms in the $CY_3$. 
Let ${\cal O} = \Omega_p (-1)^{F_L} \sigma$ be the $Z_2$ orientifold symmetry, with $\sigma$ being an antiholomorphic involution of the 
Calabi-Yau manifold. The space of $(1,1)$ forms splits into $H^{1,1}=H_{-}^{1,1}+H_{+}^{1,1}$ forms which are odd  and even under $\sigma$, 
with dimension $h_-^{1,1}, h_+^{1,1}$ respectively. Only the moduli coming from $H_{-}$ survive the orientifold projection 
\footnote{This is easy to see from their partner $B_2$ moduli which are odd under ${\cal O}$}. 
Let $J_c=B_2+i J$ be the complexified Kahler two-form.  Then we can expand $J_c$ in terms of a basis of odd two-forms, 
\beq
\label{expck}
J_c=\sum_{a=1}^{h_-^{1,1}}t_a\omega^a,
\eeq
Here,
\beq
\label{defta}
t_a = b_a+ i v_a.
\eeq
are the complexified Kahler moduli. 
The Kahler potential is given by, 
\beq
\label{kp}
K^K(t_a)=-\log({4\over 3} \int J\wedge J \wedge J) = -\log({4\over 3} \kappa_{abc}v_av_bv_c),
\eeq
where the $\kappa_{abc}$ are the triple intersection numbers, 
\beq
\label{defkabc}
\kappa_{abc}=\int \omega_a\wedge \omega_b\wedge \omega_c.
\eeq

In general there are also moduli which arise from complex structure deformations and related axions. We do not discuss them in detail here,
since these are absent in the compactification being considered in this paper. 

An single axion arises from $C_3$ in the compactification of interest to us, 
\beq
\label{c3ax}
C_3=\xi \alpha_0,
\eeq
where $\alpha_0$ is defined as the real part of $\Omega$ in eq.(\ref{reim}). 
Susy pairs this with the $4-$ dimensional dilaton defined by, 
\beq
\label{defD}
e^{2D} ={e^{2 \phi} \over vol}.
\eeq
The resulting Kahler potential for  this modulus is 
\beq
\label{dilax}
K^D=4 D. 
\eeq

Let $\tilde{w}_a$ be  a basis for $H_+^{2,2}$. These are  are  dual to the  $(1,1)$ forms  $\omega_a$ which are  a basis of $H_{-}^{1,1}$
eq.(\ref{expck}). 
 Then since $F_4$ is even under $\cal{O}$ we can expand it in this basis, 
\beq
\label{fformgen}
F_4=e_a \tilde{w}_a.
\eeq
The full superpotential has two terms in it,
\beq
\label{superpot}
W=W^Q+W^K,
\eeq
with
\beq
\label{first}
W^Q=-p\xi - \sqrt{2}i p e^{-D},
\eeq
and 
\beq
\label{second}
W^K=e_at_a-{m_0\over 6} \kappa_{abc} t_at_bt_c.
\eeq
For future use, we mention below that including  $e_0$ units of   $F_4$ flux  turned on along the non-compact directions (or 
equivalently dual $F_6$ flux  turned on along all compact directions), and   
 $m_a$  units of  $F_2$ flux turned on along compact two -cycles gives the more general superpotential for Kahler moduli,
\beq
\label{fulltwo}
W^K=e_0 + e_a t_a +{1\over 2} \kappa_{abc} m_a t_b t_c -{m_0\over 6} \kappa_{abc} t_a t_b t_c,
\eeq
while keeping $W^Q$ unchanged. 

We now restrict restrict ourself to the orbifold limit, working with only the untwisted Kahler moduli  and the
the $F_4$ fluxes eq.(\ref{fourform}). 
There are three complexified Kahler moduli from the untwisted sector, 
\beq
\label{csfa}
t_i=b_i+i v_i, i= 1, \cdots 3.
\eeq
The $v_i$ were introduced above in eq.(\ref{kmoduli}). The $b_i$ arise from $B_2$. 
\beq
B_2=\sum_{i=1}^3 b_i \omega_i.
\eeq
The triple intersection number on $T^6/(Z_3)^2$ is given by
\beq
\label{triint}
\kappa_{123}=\int_{T^6/(Z_3)^2}\omega_1 \wedge \omega_2\wedge \omega_3 = \kappa.  
\eeq
The
  resulting Kahler potential for these moduli  is then \footnote{The Kahler potential for the moduli which survive
after the orientifolding is done, is inherited from the parent ${\cal N}=2$ theory.} 
\beq
\label{kpmod}
K^K=-\log(8 \kappa v_1v_2v_3). 
\eeq

Now with the superpotential, eq.(\ref{first}), eq.(\ref{second}), and the Kahler potential eq.(\ref{dilax}),  eq.(\ref{kpmod}), 
 one gets a  potential
\begin{eqnarray} 
\label{fullpotential}
\nonumber
 V &=& {e^{2D}  p^2 \over 4 vol} + {m_0^2 vol e^{4D} \over 2}  + {e_i}^2 {v_i}^2 { e^{4D} \over 2 vol} + \sqrt{2} e^{3D} m_0 p + e^{4D} {\left( b_i e_i - p \xi \right)^2 \over 2 vol} \\
\nonumber
&&+ {m_0^2 vol e^{4D} \over 2} \left( {b_1^2 \over v_1^2} + {b_2^2 \over v_2^2} +{b_3^2 \over v_3^2}   \right) - { m_0 e^{4D} \kappa b_1 b_2 b_3 \over vol} \left( {e_1 v_1^2 \over b_1}+{e_2 v_2^2 \over b_2}+{e_3 v_3^2 \over b_3}\right)\\
\nonumber
&&-{m_0 \kappa b_1 b_2 b_3 e^{4D}\over vol} \left( b_i e_i - p \xi  \right)  + {m_0^2 (\kappa b_1 b_2 b_3)^2 e^{4D} \over 2 vol} \left( {v_1^2 \over b_1^2} + {v_2^2 \over b_2^2} +{v_3^2 \over b_3^2} \right) \\
&&+ { m_0^2 e^{4D} (\kappa b_1 b_2 b_3)^2 \over 2 vol} .
\end{eqnarray}
Setting all the axions, $b_i, \xi=0$ it is easy to see gives  the potential in eq.(\ref{potential}).
Minimizing eq.(\ref{potential}) gives,
\beq
\label{valvi}
v_i={1\over |e_i|}\sqrt{{5\over 3}|{e_1e_2e_3\over\kappa m_0}|},
\eeq
\beq
\label{vald}
e^D=|p|\sqrt{{27\over160}|{\kappa m_0\over e_1e_2e_3}|},
\eeq
and,
\beq
\label{valphi}
e^{\phi}={3 \over 4} |p| \left( {5 \over 12} {\kappa \over |m_0 e_1 e_2 e_3| } \right)^{1 \over 4}.
\eeq
The  potential at the minimum takes the value,
\begin{equation}
\label{potmin}
V_0=-{2 |e_1e_2e_3|\over 3 \kappa v} e^{4 D}.
\end{equation}
where $v_i={v\over |e_i|}$. In terms of fluxes this is,
\begin{equation}\label{potpar}
 V_0 = - \sqrt{4 \over 15} \left( 27 \over 160 \right)^2 {p^4 \kappa^{3 \over 2} |m_0|^{5 \over 2} \over |e_1 e_2 e_3|^{3 \over 2}}.
\end{equation}
Keeping the terms in action   which are quadratic in the axions gives, 
\begin{eqnarray}
\label{actax}
\nonumber
S_{axion} & = & {1\over 2 } \int d^4x\sqrt{-g}\bigl(-\sum_i {1\over 2} \partial_\mu \tilde{b}_i \partial^\mu \tilde{b}_i   - e^{4 D} (m_0^2  vol \ \tilde{b}_i^2 - 2 m_0 \tilde{b}_1 \tilde{b}_2 \tilde{b}_3 {e_i v_i \over \tilde{b}_i})\\
&& - {1\over 2} \partial_\mu x \partial^\mu x -{e^{4D}\over vol}(\tilde{b}_1e_1v_1+ \tilde{b}_2e_2v_2+\tilde{b}_3 e_3 v_3-{p\over \sqrt{2}} e^{-D}x )^2) \bigr).
\end{eqnarray}
where $\tilde{b}_i={b_i \over v_i}, x=\sqrt{2} e^D \xi$ are the canonically normalized axion fields. 
The resulting mass matrix for    the axions is then,  
\begin{equation}\label{massmatrix}
 M^2_{ij} = 
2 |m_{o}| e^{4 D} v \left(
\begin{array}{cccc}
{34 \over 15}   				 & {3 \over 5} s_{1} s_{2} - s_{3}     &    {3 \over 5} s_{1} s_{3} - s_{2} & {4 \over 5} s_{1} \\
{3 \over 5} s_{1} s_{2} - s_{3} 	 &  {34 \over 15}   				 &   	 {3 \over 5} s_{2} s_{3} - s_{1} &    		{4 \over 5} s_{2}			           \\ 
{3 \over 5} s_{1} s_{3} - s_{2}   &     {3 \over 5} s_{2} s_{3} - s_{1} &  {34 \over 15}  				 &  		{4 \over 5} s_{3  }			     \\
 		{4 \over 5} s_{1 }	        & 		{4 \over 5} s_{2 }	               & 		{4 \over 5} s_{3 }			 & {16   \over 15}       \\
\end{array}
\right)
\end{equation}
with $i,j, = 1,\cdots 3$ being the $\tilde{b}_i$ directions, and $i,j=4$ being $x$, and where $s_i=\text{sign}(m_0 e_i)$. 

There are two distinct cases which arise for the eigenvalues of the mass matrix. 

\smallskip
\noindent
$\bullet$ 
When $\text{sign}(m_0 e_1e_2e_3)=-1$ it turns out that  all eigenvalues are  positive. This includes the susy case where $\text{sign}(m_0e_i)=-1$ for each value of $i$.
But it also includes non-susy cases where the condition eq.(\ref{condb}), 
 is not met and  the condition $\text{sign}(m_0 e_1 e_2 e_3)=-1$  still holds.

\smallskip
\noindent
$\bullet$
When $\text{sign}(m_0 e_1 e_2 e_3)=+1$ and susy is necessarily broken, there is one negative eigenvalue and thus one  tachyon.  
Its mass is given by, $M^2=-{4\over 15}|m_0| e^{4D} v$. The BF bound is $M_{BF}^2=-{3\over 4} V_{min}$.
From eq.(\ref{potmin}), we see that 
\beq
\label{bfboundcomp}
{M^2 \over M^2_{BF}}={-({4\over 15}) |m_0|e^{4D}v \over -({3\over 10})|m_0|e^{4D}v }={8\over 9}.
\eeq
Thus we see that the mass lies above the  BF bound and hence the resulting vacua are stable with respect to these axionic directions. 

\subsubsection{Some General Comments}

Some important  features of the vacua which arise from eq.(\ref{valvi})- eq.(\ref{potpar})  are  worth emphasizing at this stage. For 
the purpose of scalings, in this section we work in string frame with string scale set equal to one. We will be 
 interested in vacua where the four-form
flux $|e_i| \sim e \gg 1$. 
Note that the tadpole condition eq.(\ref{tadpole}) imposes no constraint on the four form flux $e_i$, which is allowed to get arbitrarily large.
From eq.(\ref{valphi}) we see that the dilaton $e^\phi\sim e^{-3/4}\rightarrow 0$, so that for large flux one has a weakly 
coupled theory.

For $ e \gg 1$  we find parametrically that $\gamma_i \sim e^{1/2}$, so that the size of the internal space  scales like
$l \sim e^{1/4}$. In contrast the potential scales like $e^{3/2}$ and $M_{\text{pl}}^{(4)}\sim e^{3/ 2}$
 so that the radius of AdS space goes like $R_{AdS} \sim e^{3/4}$.  Thus we find that both $l, R_{AdS}$  
become parametrically large as $e\gg 1$. As a result higher derivative corrections in the $\alpha'$ expansion will be suppressed. 
This still leaves corrections which involve  higher powers of the field strengths without additional derivatives. 
The biggest worry are terms involving higher powers of $F_4$. It is easy to see that the leading term which goes like $F_4^2$ scales like 
$O(1)$, with two powers of $e$ coming from $F_4$ and four powers of $e^{-1/2}$ coming from the inverse metric. 
The correction due to  $F_4^4$ terms  are then    also of $O(1)$. However, as noted in \cite{DGKT}
 these are suppressed by an additional power of $g_s^2 \sim e^{-3/2}$ making them subdominant.     

Note that the ratio $l/R_{AdS} \sim {1 \over \sqrt{e}} \rightarrow 0$, so that the compactification is not of Freund-Rubin type.
The non Freund-Rubin nature of the compactification 
actually simplifies the analysis when it come to
checking for possible tachyons. The KK modes have positive $(mass)^2$ in the absence of flux, including the effects
of flux cannot make them tachyonic because of the parametric separation of scales. Thus it is sufficient to look for possible tachyons among the moduli,
which are massless in the absence of flux.

In the discussion above we saw that  the $F_4^2$ terms  scale like $O(1)$ in string units. 
In susy breaking vacua the flux sets the scale of 
susy breaking and one might therefore worry that our starting point, which is a Calabi-Yau orientifold with supersymmetry, is itself 
inconsistent. However this is not true. The gravitational backreaction of the flux  is parametrically  suppressed
for large $e$, as we argued in the previous paragraph since $l/R_{AdS} \rightarrow 0$. 
 From a ten dimensional point of view this follows from the fact that
   the gravitational  back reaction is suppressed by $g_s^2$ which is small. 
Corrections to the Calabi-Yau metric can be systematically calculated in an expansion in $1/e$. Such corrections arise in the supersymmetric
case as well and  including them in the susy case  alters the internal metric so that it  is no longer 
Calabi-Yau but instead is a half-flat metric with $SU(3)$ structure. For some discussion of this see,
\cite{CS}, \cite{Acharya}.

%



\subsubsection{Conventions}
Since numerical factors will be sometime important in the following discussion it is worth discussing our conventions in some detail. 

We  essentially follow the conventions of   \cite{DGKT}. 
More specifically, the volume of the internal space after the additional $Z_2$ orientifolding is given by ${vol\over 2}$.
To go to $4$ dimensional Einstein frame from the $10$ dim string frame  we use the relation 
\beq
\label{relframes}
g_{\mu\nu}={e^{2\phi} \over vol} g_{\mu\nu}^E.
\eeq
In addition  we set $ 2 \kappa_{10}^2=1$.
This gives the E-H term  and potential term to be
\beq
\label{EHfeb}
S= \int d^4x \sqrt{-g_E} ({1\over 2}R - V).
\eeq
where $V$ is given in eq.(\ref{fullpotential}).
The potential term  above agrees with eq.(3.21) of \cite{DGKT} after we set $\kappa_{10}^2=1$ in their formula. 
Similarly other formulae in \cite{DGKT} can be related to ours after setting $\kappa_{10}^2=1$ in those formulae.

\subsection{The Blow-up Modes}
So far we have ignored the blow-up modes. We now include them and check if there are any unacceptable tachyons which arise from the blow-up moduli
of their axionic partners.  It is convenient for us to seek vacua where the Calabi-Yau manifold is close to its orbifold limit \footnote{ 
A full analysis considering vacua far away from the orbifold point is left for the future.}. 
There are $9$ blow-up modes, turning on a blow-up mode 
 replaces the corresponding singularities with a $P^2$ of non-vanishing size. 
We will  introduce additional  $F_4$ flux threading each of these $P^2$'s.  

The complexified Kahler two-form is now given by\footnote{In our notation the index $A$ which denotes the basis elements of $H_{-}^{1,1}$ takes 
values $i =1,2,3, a=1, \cdots 9$.}. 
\beq
\label{cktwo}
J_c=\sum_a t_a \omega_a= \sum_{i=1}^3 t_i \omega_i + \sum_{A=1}^9 t_A \omega_A,
\eeq
with
\beq
\label{deftA}
t_A=b_A+iv_A.
\eeq
$\omega_A$ are elements of $H^{1,1}$ dual to the blow-up two cycles, and  $v_A, b_A$ are  the blow- up moduli and  corresponding axions. 
  
The Kahler potential in the Kahler moduli space is\footnote{$\kappa, \beta$ take values, $\kappa=81, \beta=9$, according to \cite{DGKT}, but
 we will not need
these explicit values below.}.  
\beq
\label{kpkm2}
K^K=-\log(8 \kappa v_1v_2v_3 +{4\over 3}\beta \sum_{A =1}^{9} v_A^3 ).
\eeq
Let us note that  to stay within the Kahler cone, $v_A<0$, \cite{DGKT}. 
This agrees with the intuition that   as $v_A$ increases the total volume,  which is the argument of the logarithm in the Kahler potential, decreases. 

The four- form flux is,
\beq
\label{fform2}
F_4=e_i \tilde{\omega_i} + e_A \tilde{\omega}_A.
\eeq
The superpotential is 
\beq
\label{superpot2}
W=-p \xi-i \sqrt{2}p e^{-D} + e_it_i + e_At_A-m_0(\kappa t_1t_2t_3+ {\beta\over 6} \sum_A t_A^3).
\eeq

To keep the blow-up modes small and the Calabi-Yau close to the orbifold point, we  take the extra flux $e_A$ to satisfy the 
condition, 
\beq
\label{condbflux}
{|e_A|\over |e|} \ll 1,
\eeq
where $e\sim e_i$ denotes a generic flux  along the $T^2\times T^2$ four-cycles. 
We will see below that  the resulting expectation value for 
\beq
\label{expvA}
v_A\sim \sqrt{|{e_A\over m_0}|}.
\eeq
As a result the ratio  
\beq
\label{ratbnb}
{v_A\over v_i} \sim \sqrt{|e_A|\over |e_i|} \ll 1,
\eeq
so that the blow up moduli  have a comparatively small value and the Calabi-Yau moduli will then be stabilized close to the orbifold point. 

For discussion below it is convenient to introduce the variable,
\beq
\label{defdelta}
\delta = \sqrt{|{e_A^3\over e_1e_2e_3}|}.
\eeq
To calculate the value of the blow-up modes it is enough to expand the potential and keep only the first two terms in an expansion in 
$\delta$. Keeping in mind the expected relation, eq.(\ref{ratbnb}), this means keeping terms upto order $({v_A\over v_i})^3$. 

We saw above that the leading order potential gives an acceptable extremum as far as the untwisted moduli are concerned.
The Kahler moduli and dilaton are given by, eq.(\ref{valvi}), eq.(\ref{valphi}), and the axions are zero. 
All masses lie above the BF bound. 
The first corrections will have a small effect on the untwisted moduli masses and they will continue to be safely above the
BF bound. Also, the shifts in the values at the extremum for the  Kahler moduli and the dilaton due to the first order correction
can be ignored at this order. In contrast, for the blow-up modes and their axions the first corrections 
provide the dominant potential.
In the analysis below we set   the untwisted Kahler moduli and dilaton to their minimum values, eq.(\ref{valvi}),
eq.(\ref{valphi}), and examine the effects 
of the first order corrections on the blow-up modes and their axionic partners.

First let us  set the axions to be all zero. 
This gives,
\begin{eqnarray}
\label{valpot22}\nonumber
V &=&  V_0 + \sqrt{15 \kappa \over 4 \beta} |V_0|     \sum_A \sqrt{  |e_A|^3\over |e_1e_2e_3|}  \left\lbrace 
 {1\over 2 x_A} -{3\over 10}  \text{sign}(e_A) \sum_i \text{sign}(e_i) x_A  \right. \\
&& \left. 
+{3x_A^3\over 200} (1+2\sum_{<ij>}\text{sign} (e_ie_j)  \right\rbrace,
\end{eqnarray}
where
\beq
\label{defxA}
x_A=-\sqrt{\beta |m_0|\over |e_A|} v_A.
\eeq

We note that an explicit minus sign has been introduced in the definition of $x_A$ since the 
allowed values of $v_A<0$. Also in 
the last  summation on the rhs of eq.(\ref{valpot22}) the indices $i,j=1, \cdots 3$ must take different values, and each distinct  pair $<ij>$ appears once.

Let us now introduce the axion dependence in the potential which has so far been suppressed. 
From the superpotential, eq.(\ref{superpot2})
 it follows that if  the sign of all the 
axions is reversed, keeping the Kahler moduli and the dilaton the same, then $W\rightarrow -\bar{W}$ and 
therefore the potential is invariant. This means that the first term in a power series in  the axions must be quadratic and therefore 
the extremum we find by setting them to zero is also an extremum once their dependence is included. We will examine whether this extremum is free of 
 tachyonic modes lying below the BF bound in the following discussion.  
As was mentioned above, the leading term in the potential already provides an acceptable extremum as far as the untwisted axions are concerned.  

The quadratic terms for the blow-up axions in the potential are, 
\begin{eqnarray}
\label{quadba}
\nonumber
V_{b_A} &=&{1 \over 40} \sqrt{15 \kappa \over 4 \beta} |V_0| \sum_A  \sqrt{|e_A|^3\over |e_1e_2e_3|} {b_A^2 \over (|e_A|/\beta |m_0| )} \left\lbrace-{20  \text{sign}(m_0e_A)\over x_A} \right. \\
 && +\left.  x_A ( 29+9\sum_i \text{sign}(m_0e_i) ) \right\rbrace.
\end{eqnarray}

For a solution to preserve supersymmetry eq.(\ref{condb}) must hold. In addition, since the fluxes $e_A$ are also now turned on we have the conditions,
\beq
\label{condc}
\text{sign}(m_0e_A)<0. 
\eeq

We are now ready to discuss the different cases which arise when the signs of various fluxes are varied. 
The relative sign between $m_0,p$ is fixed by the tadpole condition eq.(\ref{tadpole}). The different case will arise as we change the relative signs between
$m_0$ and the fluxes $e_0,e_A$. 
As was explained above,  our  
 main concern  as we scan over various choices of signs for the fluxes  are the blow-up modes and
their axionic partners.

\smallskip
\noindent
$\bullet$ Case 1):  $\text{sign}(e_i m_0)= \text{sign}(e_A m_0)=-1$. 
This case preserves susy. The potential has a minimum when  the blow up modes take the value,   
\beq
\label{valcase1}
x_A=\sqrt{10\over 3}. 
\eeq
All axions are  non-tachyonic including the $b_A$ axions. 

\smallskip
\noindent
$\bullet$
Case 2): $\text{sign}(m_0e_i)=\text{sign}(m_0e_A)=+1$.
Susy is broken. The extremum of the potential  
is at the same value, eq.(\ref{valcase1}). In this case there are no tachyons from the blow-up modes including the $x_A$ directions and the blow-up axions.

\smallskip
\noindent
$\bullet$
Case 3):$\text{sign}(m_0e_i)=+1$, $\text{sign}(m_0e_A)=-1$
Susy is broken, now the extremum of the potential lies at,
\beq
\label{valcase3}
x_A=\sqrt{10\over 21}.
\eeq
Again there are no tachyons from the blow-up modes and blow up axions.

Before proceeding let us note that there are $9$ blow up modes. From eq.(\ref{valpot22}), eq.(\ref{quadba})
 we see that the potential for the blow-up modes and their axionic partners
 decouple to leading order in 
$({e_A\over e_i})^{3/2}$ from each other. Thus with  $\text{sign}(m_0e_i)=1$  there are actually $2^9$ cases with  $\text{sign}(e_Am_0)$ for $A=1, \cdots 9$
 being $\pm 1$.   Depending on the sign the discussion of Case 2) or Case 3) applies for each blow-up mode and its axionic partner independently of the 
others. 

\smallskip
\noindent
$\bullet$
Case 4): $\text{sign}(m_0e_i)=-1$, $\text{sign}(m_0e_A)=+1$.
Susy is broken.  Minimum is at the same value for $x_A$ as in Case 3).
Now the $b_A$  axion is tachyonic with a mass below the BF bound, see eq. (\ref{relpp}) of appendix A. 

\smallskip
\noindent
$\bullet$
Case 5): $\text{sign}(m_0e_1)=\text{sign}(m_0e_2)=-\text{sign}(m_0e_3)=1$, $\text{sign}(m_0e_A)=+1$.
There are three cases of this type with $e_3$ being exchanged with $e_1,e_2$. 
In this cases there is no extremum for the blow-up mode at small values where the condition $|{v_A^3\over v_i^3}|\ll 1$ is met.
We have not looked for minima away from this region in moduli space.    

\smallskip
\noindent
$\bullet$
 Case 6): $\text{sign}(m_0e_1)=\text{sign}(m_0e_2)=-\text{sign}(m_0e_3)=-1$, $\text{sign}(m_0e_A)=-1$.
Again there are three cases of this type, with $e_3$ exchanged with $e_1,e_2$. And once again there is no extremum close to the orbifold limit.
  
\smallskip
\noindent
$\bullet$
Case 7): $\text{sign}(m_0e_1)=\text{sign}(m_0e_2)=-\text{sign}(m_0e_3)=1$, $\text{sign}(m_0e_A)=-1$. 
There are three cases of this type obtained by permuting the $e_i$'s. These have a point of inflection at
\beq
\label{valcase7}
x_{A*}=\sqrt{10\over 3}.
\eeq
The first two derivatives with respect to $v_A$ vanish at this point and the potential locally has the form $V\sim (x_A-x_{A*})^3$. 
To determine whether this extremum is stable we now need to calculate the next correction in the potential to order $({e_A\over e_i})^3$.
Depending on the sign of the resulting corrections a minima will arise close to the value eq.(\ref{valcase7}) or not. 
We have not carried out this calculation. 

The $b_A$ axion in this case has a positive mass at the point, eq. (\ref{valcase7}).  So if the minimum  for the $v_A$ modulus 
does lie close to this point the axion will also be  non-tachyonic.   

\smallskip
\noindent
$\bullet$
Case 8):
$\text{sign}(m_0e_1)=\text{sign}(m_0e_2)=-\text{sign}(m_0e_3)=-1$, $\text{sign}(m_0e_A)=+1$
This case is very similar to case 7). There is a point of inflection again at eq.(\ref{valcase7}). And the axion $b_A$ has positive mass.

We close this section by noting that the ground state energy for case 2),  where $(m_0e_i)>0, (m_0e_A)>0$, is given by,
\beq
\label{gndse2}
V=V_0 + V_0 \sqrt{2\kappa \over  \beta} \sum_A {|e_A|^{3/2}\over \sqrt{|e_1e_2e_3|}},
\eeq
and for case 3), where $(m_0e_A)<0, (m_0 e_i)>0$ by, 
\beq
\label{gndse3}
V=V_0-V_0\sqrt{50\kappa\over 7 \beta}\sum_A{(|e_A|)^{3/2}\over \sqrt{|e_1 e_2e_3|}}.
\eeq
These results for the ground state energy are correct \footnote{By case 2) and 3) we mean here cases where all $e_A>0$ or $e_A<0$
respectively. In the mixed case where some $e_A$ are positive and negative the terms within the sum  in eq.(\ref{gndse2}), eq.(\ref{gndse3})
 have to be changed appropriately.}
    to order $\delta$. We have neglected the shift in the untwisted Kahler moduli 
due to the blow-up fluxes, this contributes at order $\delta^2$, since the first corrections in the energy 
about the minimum are second order in the values of the moduli shifts. 

Further details leading to the  results presented above are in Appendix A, which also contains Table 1 which summarizes the different cases. 

\subsection{More General Fluxes}
In this subsection we consider what happens when $F_2$ and $F_6$ flux are also activated. 
The fluxes are specified, in terms of the basis of two-forms $\omega_a$ by, 
\beq
\label{def26f}
e_0=\int F_6, \ \ F_2=-m_a\omega_a.
\eeq
After they are turned on, $W^Q$ remains unchanged, eq.(\ref{first}).
$W^K$ now becomes, eq.(\ref{second}). 
The full superpotential is therefore,
\beq
\label{fullsupjan}
W=e_0+e_at_a+{1\over 2} \kappa_{abc}m_at_bt_c-{m_0\over 6} \kappa_{abc} t_at_bt_c-p\xi-\sqrt{2}ipe^{-D}.
\eeq
The third term on rhs, which is quadratic in $t_a$  contains the effects of the $m_a$ flux. By shifting $t_a$ by 
\beq
\label{shiftt}
t_a\rightarrow t_a-{m_a\over m_0},
\eeq
and $\xi$ by 
\beq
\label{shiftxi}
\xi\rightarrow \xi-{e_0\over p} -{e_am_a\over p}-{1\over 3} \kappa_{abc}{m_am_bm_c\over m_0^2p},
\eeq
one can can reexpress the superpotential in terms of the shifted variables  as,  
\beq
\label{reexpsu}
W={\hat e}_at_a-{m_0\over 6} \kappa_{abc} t_at_bt_c-p\xi-\sqrt{2}ipe^{-D}
\eeq
with 
\beq
\label{defhate}
\hat{e}_a=e_a+{\kappa_{abc}\over 2} {m_bm_c\over m_0}.
\eeq 
Notice that $m_a$ and $e_0$ have both disappeared in this superpotential which is the same as the sum of eq.(\ref{first}) and eq.(\ref{second}). 
The shift, eq.(\ref{shiftt}), eq.(\ref{shiftxi}), changes the real part of the chiral superfields and thus does not change the
 Kahler potential 
which is expressed in terms of the imaginary part of the   chiral superfields,  eq.(\ref{dilax}), eq.(\ref{kpkm2}).
 Thus we see that  the theory can be mapped into the one we had studied earlier,  without any $m_a$ and $e_0$ flux.

The solutions in terms of the shifted variables then are the same as those in eq.(\ref{valvi}) - eq.(\ref{potpar}),
and similar equations with the blowup fluxes also activated in \S2.3,  with $e_a$ replaced by $\hat{e}_a$. 

The conditions for supersymmetry also, it turns out, can be obtained by replacing $e_a$ with $\hat{e}_a$ and are given by 
\beq
\label{condsusy}
\text{sign}(m_0 p)<0, \ \ \text{sign}(m_0 \hat{e}_a)<0.
\eeq

The discussion in the previous subsections now carries over and we see that non-susy
 solutions  of Type 2) and Type 3) can be constructed with 
$\text{sign}(m_0 \hat{e}_a)>0$ in the first case and $\text{sign}(m_0 \hat{e}_i)<0, \ \text{sign}(m_0\hat{e}_A)>0$ in the second. 
These have vacuum energies gives in eq.(\ref{gndse2}) and eq.(\ref{gndse3}) respectively with $e_a\rightarrow \hat{e_a}$. 

One final comment is worth making regarding this case. The compactification has gauge symmetries under which the flux and moduli transform,
these can be thought of as   the analogue of the $ \tau \rightarrow \tau+1$ subgroup of $SL(2,Z)$ which arises on a torus \cite{DGKT}.
Under this symmetry, 
Configurations related by these symmetries are not distinct but should be identified. 
One set of such symmetries involve the shift in the $b_a$ axions by integer units,
\begin{eqnarray}
t_a & \rightarrow &  t_a-u_a ,\\
m_a & \rightarrow & m_a-m_0 u_a ,\\
e_a & \rightarrow & e_a +\kappa_{abc}m_b u_c-{m_0\over 2}\kappa_{abc}u_bu_c ,\\
e_0 & \rightarrow & e_0+{1\over 2} \kappa_{abc} m_au_bu_c-{m_0\over 6}\kappa_{abc}u_au_bu_c+e_au_a,
\end{eqnarray}
for integer $u_a$. The other involves the shift in $\xi$ axion, 
\begin{eqnarray}
\xi& \rightarrow & \xi-u ,\\
e_0 & \rightarrow  & e_0-p u .
\end{eqnarray}
Using these any $m_a$ which is an integer multiple of $m_0$ and $e_0$ which is integer multiple of $p$ can be set to zero. 
Since the $m_a$ and $e_0$ fluxes satisfy quantization conditions, this only leaves a few physically distinct cases where these fluxes are non-vanishing.

\section{Vacuum Decay in the Thin wall Approximation}
In this section we review the classic discussion of the non-perturbative decay of an unstable vacuum in \cite{CDL}.
Consider an unstable  vacuum, called  the false vacuum, which  can decay to 
another state, the true vacuum. The decay is mediated by the nucleation of a bubble of true 
vacuum inside the false vacuum. This nucleation is a quantum tunneling  process, and gives rise to a probability for decay  
per unit volume per unit time of the form
\beq
\label{decay}
\Gamma/V=Ae^{-B/\hbar}.
\eeq
In the semi-classical approximation one seeks a solution to the Euclidean action which can interpolate between the false and true vacua. 
Given such a solution, which is called the bounce,  the coefficient in the exponent above is given by,
\beq
\label{bounce}
B=S_E-S_{False},
\eeq
where $S_E$ is the euclidean action of the bounce and $S_{False}$ is the action of the false vacuum. 

We will work in the thin wall approximation in this paper. In this approximation the bounce solution or the 
 bubble has three parts. The inside where the solution is well approximated by the true vacuum, the outside which is the false vacuum,
and the bubble wall which interpolates between the two. In the thin wall approximation, 
the thickness of this wall is much smaller  than all the other length scales in the problem. These include the radius    of the bubble and 
the radii of curvature of the inside and outside spacetimes. 
Once these conditions are met, the tension of the bubble wall  can be calculated by taking it to be a flat wall in flat space-time,
 neglecting    both the curvature of spacetime and the curvature of the bubble wall. This   simplifies the analysis considerably.

Before we proceed a  comment  about notation is in order. 
In the conventions of \cite{CDL} the EH term and potential are given by, 
\beq
\label{cdact}
S=\int d^4x \sqrt{-g}[{1\over 2\kappa} R -U].
\eeq
These can be related to our conventions eq.(\ref{EHfeb}) by seeting  $\kappa=1$ and  taking $U\rightarrow V$. 

The Euclidean metric of the bounce solution can be taken to be $S^3$ symmetric and of form,
\beq
\label{metrice}
ds^2=d\xi^2+\rho(\xi)^2(d\Omega)^2,
\eeq
where $(d\Omega)^2$ is the volume element on a unit $S^3$. 
We will be interested in decays where the inside and outside spacetime are both $AdS$.  We denote the vacuum energy of the false vacuum, 
which is outside the bubble, and the true vacuum, which is inside, as $U_+$ and $U_-$ respectively. Both are negative. 
The bubble wall lies at $\rho=\bar{\rho}$. 

The bounce action gets contributions from the three parts of the solution, the inside, the wall and the outside, 
\beq
\label{defB}
B=B_{inside}+B_{Wall}+B_{outside}.
\eeq
Since the outside region is  essentially identical  to the false vacuum, $B_{outside}=0$.
For a wall with tension $S_1$
\beq
\label{bwall}
B_{wall}=2\pi^2 {\bar \rho}^3 S_1.
\eeq
Finally the inside region contributes, 
\begin{eqnarray}
\label{binside}
B_{inside}&=& {12 \pi^2 } \left[ { \left( 1- {1 \over 3}\bar{\rho}^2 V_-\right)^{3 \over 2} -1 \over V_-} -{ \left( 1- {1 \over 3}\bar{\rho}^2 V_+\right)^{3 \over 2} 
-1 \over V_+}  \right].
\end{eqnarray}

The value of $\bar{\rho}$ can be calculated by extremizing $B$, i.e. requiring,
\beq
\label{condrho}
{dB\over d\bar{\rho}}=0.
\eeq

For small values of $\bar{\rho}$, $B_{inside}\simeq -{\pi^2\over 2} \epsilon {\bar \rho}^4$, where
\beq
\label{defeps}
\epsilon=V_+-V_-.
\eeq
is the energy difference between the two vacua. 
Therefore for small $\bar{\rho}$,
\beq
\label{smallr}
B\simeq 2 \pi^2 {\bar \rho}^3 S_1  -{\pi^2\over 2} \epsilon {\bar \rho}^4.
\eeq
This is exactly what we would get in the absence of gravity. The bulk gain in energy grows more rapidly
with increasing $\bar{\rho}$ than the surface tension and there is 
 always  a solution to the minimization condition eq.(\ref{condrho}). 

However,  now notice from eq.(\ref{binside})  that with $U_{\pm}<0$, for  large $\bar{\rho}$
\beq
\label{larger}
B_{inside}= -{4\over\sqrt{3}}\pi^2\bar{\rho}^3(\sqrt{|V_-|}-\sqrt{|V_+|}),
\eeq
so that the bulk gain in energy  only grows like $\bar{\rho}^3$ and not as $\bar{\rho}^4$. 
This is a consequence of the fact 
that the volume and the area both grow in the same fashion in $AdS$ space. The wall contribution which goes like the area
also grows like $\bar{\rho}^3$ with a positive coefficient, eq.(\ref{bwall}). 
At small $\bar{\rho}$ the leading contribution
to $B$ comes from $B_{wall}$ and is positive. Thus a sufficient condition for an extremum value of $B$ to exists is that the net 
coefficient of the $\bar{\rho}^3$ dependence at large $\bar{\rho}$ is negative. 
For $|V_-|>|V_+|$  this yields the condition, 
\beq
\label{condsuff}
S_1<\sqrt{4\over 3}[\sqrt{|V_-|}-\sqrt{|V_+|}].
\eeq
We will see below that this is also a  necessary condition  for an extremum solving eq.(\ref{condrho})
to exist.

We now proceed with a more detailed  analysis of eq.(\ref{condrho}).
From eq.(\ref{defB}), eq.(\ref{bwall}, eq.(\ref{binside}), this gives the condition, 
\begin{equation}
\label{condf}
\left( 1- {1 \over 3}\bar{\rho}^2 V_-\right)^{1 \over 2} - 
\left( 1- {1 \over 3}\bar{\rho}^2 V_+\right)^{1 \over 2} = {\bar{\rho} S_1 \over 2}.
\end{equation}
Multiplying both sides by $\left( 1- {1 \over 3}\bar{\rho}^2 V_-\right)^{1 \over 2} + \left( 1- {1 \over 3}\bar{\rho}^2 V_+\right)^{1 \over 2} $ and simplifying, we get
\begin{equation}
 \left( 1- {1 \over 3}\bar{\rho}^2 V_-\right)^{1 \over 2} + \left( 1- {1 \over 3}\bar{\rho}^2 V_+\right)^{1 \over 2} = {2 \over 3 S_1} \bar{\rho} \left( V_+ - V_-\right).
\end{equation}

Some more algebra then gives, 
\beq
\label{condga}
 2 \left( 1- {1 \over 3}\bar{\rho}^2 V_-\right)^{1 \over 2} =
 {\bar{\rho} S_1 \over 2} +  {2 \over 3 S_1} \bar{\rho} \left( V_+ - V_-\right), 
\eeq
\beq
\label{condgb}
2 \left( 1- {1 \over 3}\bar{\rho}^2 V_+\right)^{1 \over 2} =
 -{ \bar{\rho} S_1 \over 2} +  {2 \over 3 S_1} \bar{\rho} \left( V_+ - V_-\right).
\eeq

Note that   eq.(\ref{condgb}) gives the requirement that ${\epsilon \over 3 S_1} > { S_1 \over 4}$.
In particular this requires that $\epsilon>0$ so that the inside vacuum has a lower energy than the outside one, as one would expect 
physically. Solving  eq.(\ref{condgb})  we get,
\begin{equation}
\label{exprrho}
 \bar{\rho} =  {1 \over \sqrt{ { V_+ \over 3} + \left( {\epsilon \over 3 S_1} 
- {S_1 \over 4}\right)^2 } }.
\end{equation}

For a bounce solution to exist it is necessary that $\bar{\rho}$ is finite. This gives the condition 
\begin{equation}
\label{finalcond}
 \left( {\epsilon \over 3 S_1} - { S_1 \over 4} \right) > \sqrt{ |V_+| \over 3},
\end{equation}
which  can also  be expressed as 
\beq
\label{secondcond}
{S_1^2\over 4}+\sqrt{|V_+|\over 3}S_1-{\epsilon \over 3}<0.
\eeq
It is easy to see that this condition can be met iff eq.(\ref{condsuff}) is true. 

Let us end this section with a few  comments.  If the outside vacuum is flat, with $U_+=0$, there continues to be a condition on the 
tension,
\beq
\label{thirdcond}
S_1<\sqrt{4\over 3}|V_-|^{1/2},
\eeq
which must be met for the bounce solution to exist. 
Looking back at eq.(\ref{binside})
 one finds that in this case  the large $\bar{\rho}$ behaviour of $B_{inside}$ continues  to be of the form $\bar{\rho}^3$,
and is driven by the $AdS$ vacuum inside. As a result the tension must again be smaller than a critical value for the decay to proceed. 

If the outside vacuum is  deSitter, with $V_+>0$,  it follows from eq.(\ref{exprrho}) that  $\bar{\rho}$ is always finite.
For the decay to be allowed the condition $\epsilon>0$ must be met of course.  
However  there seems to be an  additional condition which arises from eq.(\ref{condgb}) that gives the requirement 
 ${\epsilon \over 3 S_1} > {S_1 \over 4}$.
In fact this requirement does not have to be met\cite{BT}. It turns out that there are additional possibilities
for the decay of dS space which arise because  Euclidean deSitter is a sphere and therefore 
a compact manifold.
Allowing for these additional possibilities 
removes this requirement.

\section{$D4$-Brane Mediated Decays}
In this section we  consider non-perturbative decays 
 mediated by a domain wall that carries only $D4$ brane charge.
More general brane configurations carrying other charges as well will be discussed later. 

The $D4$ brane   wraps a two-cycle in the internal space and 
extends along two of the $3$ spatially non-compact directions of $AdS_4$ thereby giving  rise to a domain wall
which separates the true vacuum from the false one. 
  This causes the four-form flux $F_4$,  along the $4$-cycle dual to the two-cycle wrapped by the
$D4$, to jump. And this change in $F_4$  cause a change in the  cosmological constant. 

We will work within the thin wall approximation below. 
For this approximation to hold the domain wall must have a  a thickness which is much smaller than all the other relevant length scales, 
namely, the  radius of the $S^3$,
and the AdS radii of the true and false vacua. At first sight it would seem that this condition is obviously met since 
a  $D4$-brane  is much thinner than all these 
distance scales, when supergravity is valid.
However there is an important caveat,  which was also  mentioned in the introduction.    
In the situations at hand a   change in  flux also causes the vacuum expectation value  of the moduli to change. 
As a result  the moduli also begin to  vary across the domain wall. Now the  moduli have a mass of order the $R_{AdS}^{-1}$, see Appendix B.
 Thus 
their  variation results in a wall with  thickness  of order $R_{AdS}$,  which is not thin. 

To stay within the thin wall approximation we will only consider decays where the change in the moduli from one vacuum 
to the other is sufficiently small. The  moduli  contribution to the tension will then be  much  smaller than the $D4$ brane 
contribution and can therefore be neglected.  
The  domain wall can  then  be well approximated by a   $D4$-brane which    is indeed thin. 
The precise conditions  ensuring that the moduli contribution is small   will be worked out for various cases as we proceed. 

This section is quite long. The essential arguments can be found in the beginning few subsections, \S4.1 - \S4.3 and also in \S4.8.

\subsection{Non-Susy to Susy Decay}
We begin by  considering  the  decay of a  non-susy vacuum to a susy one. 
%
We will see below that for all these decays the tension of the domain wall is  larger than the energy difference between the two vacua, resulting 
in the decays being forbidden in the thin wall approximation \footnote{More correctly the necessary condition  eq.(\ref{finalcond}) will not be met.}. 
This mismatch is parametric in the flux, therefore in this subsection we do not need to keep track of precise numerical factors. 

To begin, we work in the orbifold limit, neglecting the blow-up modes and the related fluxes, $e_A$. 
This leaves three two-cycles, namely the three $T^2$'s, and three fluxes, $e_i, i=1,2,3$. 
The essential argument will become clear if we take 
all the three fluxes $e_i$ to be of the same order, $e_i \sim e$. For the supergravity description to be valid 
 $|e| \gg 1$.

Now consider a single $D4$ brane which  wraps the first $T^2$. Its tension arises from the Nambu-Gotto action, 
\beq
\label{sd4}
{\rm Action} = \mu_4 e^{-\phi} \int \sqrt{-g}d^3\xi_i dx^1dy^1 \sim   \gamma_1 \mu_4 e^{-\phi} \int \sqrt{-g} d^3\xi_i.
\eeq
where $\xi_i, i=1, \cdots 3$ are the $3$ directions in $AdS$ space along which it extends, and we have used eq.(\ref{metric})
 and done the integral over the $T^2$.
 
We will work in $4$ dim Einstein frame below. This is related to the string frame by 
\beq
\label{ef}
g_{\mu\nu}={e^{2\phi}\over vol} g_{\mu\nu}^E.
\eeq
Accounting for this, gives the Einstein frame tension for a single $D4$ wrapping the first $T^2$ to be 
\beq
\label{eften}
S_1\sim \gamma_1 \mu_4 e^{-\phi} ({e^{2\phi}\over vol})^{3/2},
\eeq
where $\gamma_1$ is the size of the $T^2$. 

The domain wall of interest  to us is obtained by wrapping all three $T^2$'s in general, since all three fluxes must reverse sign.
Its tension is of order  
\beq
\label{torder}
S_1\sim \gamma \mu_4 e^{-\phi} ({e^{2\phi}\over vol})^{3/2} |\delta e|,
\eeq
where $\gamma\sim \gamma_i$ is the size of the 3 $T^2$'s and $\delta e\sim \delta e_i$ is the change in flux. 
Using eq.(\ref{kmoduli}) and  eq.(\ref{valvi}), eq.(\ref{valphi}), eq.(\ref{vald}), we get
\beq
\label{paratension}
S_1 \sim {|\delta e|\over |e|}  ({1 \over |e|})^{9/4},
\eeq
where $\delta e \sim \delta e_i$ is the change in the flux. 
 
Now let us take into account the conditions imposed by the  thin wall approximation. 
From Appendix B we see that the moduli make a contribution to the tension 
\beq
\label{modcontr}
T_{mod}\sim M  (\Delta \Phi)^2.
\eeq
Here $M \sim 1/R_{AdS} \sim 1/|e|^{9/4}$,  is the mass of canonically normalized moduli field, and 
\beq
\label{chngmod}
\Delta \Phi \sim {\delta v_i\over v_i} .
\eeq
is the total change 
in the vacuum expectation value of the canonically normalized field across the domain wall.
The  value  of the  $v_i$ moduli in any vacuum depend on the absolute value of the flux. This gives 
\beq
\label{condjand}
\Delta \Phi \sim {\delta |e|\over |e|}.
\eeq 
Substituting in eq. (\ref{modcontr}) yields
\beq
\label{tmod2}
T_{mod}\sim {(\delta |e|)^2 \over |e|^2} ({1 \over |e|})^{9/4}.
\eeq
For the thin wall approximation to hold,  $S_1$ must  dominate over $T_{mod}$. This gives 
\beq
\label{conddd}
{|\delta e|\over |e|}\gg {(\delta |e|)^2 \over |e|^2}.
\eeq

Now this condition can be met if the susy vacuum has fluxes which are opposite in sign but approximately the same in magnitude
as the non-susy vacuum we start with.  That is, 
\beq
\label{congg}
e^{susy}_i \sim   e^{susy} \sim - e.
\eeq 
Then the non-susy  vacuum we start with and the susy vacuum it could decay to, 
 lie in approximately the same region of 
moduli space, but are very far apart in flux space. As a result 
\beq
\label{rel4a}
|\delta e| \sim  2 |e|,
\eeq
 and eq.(\ref{conddd}) becomes,
\beq
\label{condde}
1\gg {(\delta |e|)^2 \over |e|^2}.
\eeq
From eq.(\ref{congg}), $|e^{susy}|\sim |e|$  and therefore  $\delta |e| =  |e|-|e^{susy}|$ is small and this condition is indeed met 
 \footnote{We cannot take $|\delta e | \gg 2 |e|$
because then $|\delta e| \sim \delta |e|$ and thus eq.(\ref{conddd}) will not be met. 
}.  

Having ensured that the decay process lies within the thin wall approximation let us now see why it is not allowed.  
The important point is that since the absolute value of the flux in the non-susy and susy vacua are close, their energy difference is also small. 
From  eq.(\ref{potmin}), 
\beq
\label{epsd4}
\epsilon\sim {\delta|e| \over |e|} ({1 \over |e|})^{9/2}.
\eeq
For the decay to proceed,  a necessary condition which follows from eq.(\ref{finalcond})
 is that ${\epsilon\over 3 S_1}>{ S_1\over 4}$. 
 From, eq.(\ref{paratension}) and eq.(\ref{epsd4}),  this condition becomes, 
\beq
\label{condd43}
{\delta|e|\over |e|} \ge ({|\delta e|\over |e|})^2 \sim O(1),
\eeq
where the last relation  follows from eq.(\ref{rel4a}).   We see now that the condition in eq.(\ref{condd43})
is incompatible with eq.(\ref{condde}). 

Thus we see that the  decay of  a non-susy vacuum to a susy vacuum is not allowed in the thin wall 
approximation. 

So far we have neglected the blow-up modes and also neglected the related blow-up fluxes. After including these one can have perturbatively stable 
 non-susy
 vacua of  Type 2) or Type 3) as discussed in the \S2. The obstruction we found above  disallowing  a non-susy to susy decay  was 
parametric in the $e_i$ fluxes for large $|e_i|$. 
Including the blow-up fluxes cannot overcome  this parametric obstruction  as long as  the blow-up fluxes are small and meet the condition,
 eq.(\ref{condbflux}). 
Therefore  we conclude that non-susy vacua of Type 2) and 3), which arise when the flux meets the condition  eq.(\ref{condbflux}),  
cannot decay to susy vacua in the thin wall approximation.


\subsection{Decays From Non-Susy to Other Non-Susy Vacua}\label{sec4.2}
We now turn to examining whether a non-susy vacuum can decay to other non-susy vacua. 
We will need to calculate the tension of a $D4$ brane wrapping a two-cycle in the internal space. 
The $D4$-brane causes a jump in the flux to occur and therefore a jump in the superpotential. 
It is well known that the tension of the resulting domain wall  satisfies a lower bound 
\beq
\label{lbfeb}
T\ge T_L,
\eeq
 where $T_L$ is   given by 
\beq
\label{lb2jan}
T_L= 2  e^{K/2} |\Delta W|,
\eeq
with
\beq
\label{chngwjan}
\Delta W = \delta e_a v_a,
\eeq
 being  the change in the superpotential  caused by the jump in the flux.

Our basic strategy will be to compare $T_L$ with an upper bound in terms of the energy difference between the two vacua. 
This  will allow us to rule out various decays.  

Let us note here that the formula, eq.(\ref{lb2jan}) is  true more generally as well, when the $D$ brane carries other charges too. 
It arises because the tension of the domain wall 
is only determined by the geometry of the Calabi-Yau space, in our approximations. 
In fact, in the absence of fluxes,
the Calabi-Yau manifold preserves supersymmetry and the lower bound for the tension, in terms of the jump in the superpotential, is really
 a BPS bound. Branes which saturate the bound preserve supersymmetry, in the absence of flux.

In the $D4$-brane case the lower bound  follows  from the fact that    the Kahler form on the Calabi-Yau is a calibration \footnote{For an early reference see
\cite{GPT}. For a pedagogical discussion see, \cite{Gubser}.}.
For the sake of clarity let us pause to quickly review how this comes about.  This will also allow us to fix the normalization constant in eq.(\ref{lb2jan}).  

Consider a two- cycle in the Calabi-Yau manifold. Let $\sigma, \bar{\sigma}$ be holomorphic and anti-holomorphic
coordinates on the world volume. And let the Pull back of the Kahler form of the Calabi-Yau onto the world volume be, 
\beq
\label{pbk}
P[J]=K_{\sigma\bar{\sigma}}d\sigma d\bar{\sigma}.
\eeq
Also let $P[g]$ be the induced metric on the world volume (we are supressing indices here). The area element is then given by,
\beq
\label{aelemnt}
\sqrt{det(P[g])} d\sigma \wedge d\bar{\sigma}.
\eeq
Now since the Kahler form of the Calabi-Yau is a calibration we know that for any two-cycle, 
\beq
\label{ineq}
|K_{\sigma\bar{\sigma}}|\le \sqrt{det(P[g])}.
\eeq
The equality is met only when the cycle is either holomorphic, or antiholomorphic. In the holomorphic case
$z^i(\sigma)$ where $z^i$ are coordinates of the Calabi-Yau manifold; in the antiholomorphic case, $z^i({\bar \sigma})$.
In these cases the $D4$ brane wrapping the two-cycle is supersymmetric \footnote{For a more extensive discussion see \S4.9.}.

The tension of the resulting domain wall is given by 
\beq
\label{tensionjan}
T=\mu_4 e^{-\phi} \int d^2\sigma \sqrt{det(P[g])} .
\eeq
Using eq.(\ref{ineq}) we get a lower bound on the tension, 
\beq
\label{lbt}
T\ge T_L \equiv\mu_4 e^{-\phi}|\int P[J]|.
\eeq
Now if $\gamma_a$ is a basis of two-cycles and $\omega_a$ a basis of dual-two forms, and if the two-cycle wrapped by the $D4$ brane 
is $\gamma=\delta n_a\gamma_a$ then we have 
\beq
\label{intkjan}
\int P[J]=\delta n_a v_a,
\eeq
where the Kahler moduli $v_a$ are defined in eq.(\ref{expck}). 
This leads to the lower bound 
\beq
\label{lbjan}
T_L=\mu_4 e^{-\phi}|\delta n_a v_a|.
\eeq
We now go to 4 dim Einstein frame using eq.(\ref{ef}). In addition we relate the winding numbers  $\delta n_a$ 
to the jump in  the four form flux  $\delta e_a$  by
\beq
\label{relenjan}
\delta e_a=3^{2/3} \sqrt{2}\kappa^{1/3}\mu_4 \delta n_a,
\eeq
as discussed in Appendix C.  Finally we also include an extra normalization factor having to do with our definition of the Kahler two-form,
which is also discussed in Appendix C. Altogether we then get eq.(\ref{lbjan}), with eq.(\ref{chngwjan}).  

Let us now turn to evaluating the upper bound on the tension.  
We saw in \S3 that for the decay to be allowed it must meet the   condition, eq.(\ref{condsuff}).
Using the definition of $\epsilon$ in eq.(\ref{defeps}), and working to leading order in $\epsilon$ we get,
\begin{equation}\label{TU}
T \le T_U\equiv  {\epsilon \over \sqrt{3   |V_0|}}.
\end{equation}
The justification for working to leading order in $\epsilon$  comes from the thin wall approximation, as we will see below.

In the discussion below we will ask if the lower bound  $T_L$ is bigger than the upper bound $T_U$.
If this is true the decay will not be allowed. In cases where $T_L<T_U$ there will be interpolating $D4$ branes.e.g. wrapping susy
cycles which saturate the lower bound, which will lead to allowed decays.

Before proceeding let us make one more  comment. 
When we calculated $T_L$ above  we  assumed that the moduli are fixed and calculated the tension of the $D4$ brane in this fixed moduli 
background. Actually the change in flux caused by the $D4$ brane also causes the moduli to change. But as long as the fractional change in 
expectation value of the moduli is small, i.e.,  
\beq
\label{fracfc}
|{\delta v_a\over v_a}|\ll 1,
\eeq
the resulting effect on the $D4$ brane tension can be neglected.  In the thin wall approximation the variation of the moduli must make a smaller 
contribution to the domain wall tension than the $D4$-brane makes, this requirement gives  rise to the condition,
eq.(\ref{fracfc}), as we will see below.

\subsection{Decays in  the Orbifold limit}
To begin let us set the flux $e_A$ 
along the blow-up $4$-cycles to be zero. Only the $e_i$ fluxes are then activated and we only consider  $D4$ branes which
wrap the $T^2$ two-cycles and cause these fluxes to jump.

The limitations imposed by the thin wall approximation can be understood in terms of a discussion analogous to the one in \S4.1.
The important difference here is that since the true and false    vacua are non-susy of Type 2) or 3) the $e_i, i=1, \cdots 3$ fluxes 
have the same sign in both of them. To save clutter we set $m_0>0$ below.  Then $e_i>0$ in these vacua. 
To begin let us consider a case where all the fluxes are comparable, $e_1\sim e_2\sim e_3\sim e$, and where the change in flux caused
by the domain wall is also comparable, $\delta e_i \sim \delta e$. 
A  $D4$ brane which changes the flux by amount $\delta e$  contributes a tension, 
\beq
\label{ten1jan}
T_{brane}\sim |{\delta e\over e}| {1\over |e|^{9/4}}.
\eeq
The moduli contribute a tension which is now, 
\beq
\label{tmodjan}
T_{moduli}\sim |{\delta e\over e}|^2 {1\over |e|^{9/4}}.
\eeq
For $T_{brane}\gg T_{moduli}$ we get, 
\beq
\label{condjan1}
|{\delta e\over e}|\ll 1.
\eeq

From eq.(\ref{valvi}) we see that the moduli change in response to the flux by  
\beq
\label{valvjan}
{\delta v_i \over v_i}\sim {\delta e \over e}.
\eeq
Thus  eq.(\ref{fracfc}) follows from the condition, eq.(\ref{condjan1}). 
From eq.(\ref{potmin}) the potential at the minimum  changes by 
\beq
\label{potjan1}
|{\epsilon\over V_0}| \sim |{\delta e\over e}| \ll 1.
\eeq
This justifies working to leading order in $\epsilon$ in eq.(\ref{TU}).  

If the three $e_i$  fluxes and/or their changes are not comparable, a similar argument goes through with the factor $|{\delta e\over e}|$
being  replaced by that for the  flux with the largest fractional change, i.e. the largest values of $|{\delta e_i \over e_i}|$. 
Once again, both eq.(\ref{fracfc}), and eq.(\ref{TU}) follow.


We now calculate both $T_U$ and $T_L$ for such decays. The energy difference
$\epsilon$ can be calculated in terms of the change in fluxes $\delta e_i$ from eq.(\ref{potpar}). This gives, 
\begin{equation}
\label{tempjan1}
 T_U  =  {3 \over 2} (\sum_i {\delta e_i \over e_i}) \sqrt{|V_0| \over 3 }.
\end{equation}
From eq(\ref{defvol}), eq(\ref{valvi}) and eq(\ref{potmin}), we get
\begin{equation}\label{rel1}
 {\sqrt{|V_0|} \over e_i} = v_i \sqrt{2 \over  3  } {e^{2 D} \over \sqrt{vol}}.
\end{equation}
Eq.(\ref{tempjan1}) then becomes, 
\begin{equation}\label{TUuntw}
 T_U = {e^{2D} \over \sqrt{2 vol} } \sum_i \delta e_i v_i.
\end{equation}
For the decay to occur $\epsilon>0$, this means $ \sum_i \delta e_i v_i >0$.
As a result we get, 

Next we calculate $T_L$.  
Since $\Delta W = \delta e_i v_i$, and Kahler potential $K$ is as given in eq(\ref{kpmod}), using eq(\ref{valvi}), we get
\begin{equation}\label{TLuntw}
 T_L = 2 e^{K \over 2} |\Delta W| = {e^{2D} \over \sqrt{2 vol} } \sum_i \delta e_i v_i.
\end{equation}
Comparing  eq(\ref{TUuntw}) and eq(\ref{TLuntw}), we see that  $T_U=T_L$. 

This means the decay can  at best be  marginally allowed. The marginal cases arises when the $D4$ brane wraps a supersymmetric cycle.
  Since supersymmetry is broken one expects that corrections to the approximation we are working in will result in the marginal case
becoming either allowed or disallowed \footnote{The marginal case corresponds to a no-force condition on the $D4$ brane and a flat direction
in the $AdS_4$ theory. Such a flat direction should  get lifted  without susy.}. 
We will incorporate some of these corrections in the following discussion and also comment  on which cases remain marginal  after including some of these corrections 
further below.  


In the discussion above we have set the $e_A$ fluxes to vanish. If they are turned on but are small so that $\delta$ defined in 
eq.(\ref{defdelta}) is small, then for $D4$ branes which only wrap the $T^2$ two-cycles the calculations above still give the leading answers 
in $\delta$ for $T_U, T_L$.  In the discussion below we will now turn to including $D4$ branes which can cause a change in 
the  $e_A$ fluxes.  

\subsubsection{Explicit Example of a Disallowed Decay}
The advantage of working in the orbifold limit is that one can explicitly calculate the size of the $T^2$ two-cycles and associated tension of branes. 
This allows us to give a simple example of a situation where the brane tension is too big, because the cycle is not holomorphic resulting in 
 the decay being disallowed. 

Consider a $D4$ which wraps the first two $T^2$'s. Let $\sigma, \bar{\sigma}$ be the holomorphic, and antiholomorphic  coordinates on the world volume,
and let the mapping from the world volume to the first $T^2$ be linear and holomorphic,  $z^1(\sigma)=\sigma,$ and to second $T^2$  be linear and 
 anti-holomorphic, $z^2(\bar{\sigma})=\bar{\sigma}$. The resulting cycle is clearly not holomorphic, and the resulting wrapping numbers for the two $T^2$'s are $+1$ and $-1$ respectively.   
The tension of the resulting $D4$ brane is  
\beq
\label{tensionexjan}
T={e^{2D}\over \sqrt{2vol}} \delta |e_1|(v_1 + v_2).
\eeq
From the discussion above we have, 
\beq
\label{secondcomp}
|T_U|=T_L={e^{2D}\over \sqrt{2vol}} \delta |e_1||(v_1 - v_2)|.
\eeq
Thus we see that $T>T_U$ and the decay is not allowed. 

\subsection{General Decays With Blow-up Fluxes}
Let us first examine the conditions imposed by the thin wall approximation on the allowed change in the blow-up fluxes. 
From the Kahler potential eq.(\ref{kpkm2}) it is easy to see that a change in  canonically normalized blow-up modes is, 
\beq
\label{cannormb}
\Delta \phi_{bu}\sim \sqrt{\delta} {\delta v_A \over v_A},
\eeq
where $\delta v_A$ is the change in the blow-up moduli. 
It then follows that the blow up modes also have a mass, 
\beq
\label{bumass}
M_{bu}\sim \sqrt{|V_0|} \sim R^{-1}_{Ads},
\eeq
and their contribution to the tension is 
\beq
\label{cbutension}
T_{bu} \sim M_{bu} \Delta (\phi_{bu})^2 \sim \sqrt{|V_0|} \delta  ({\delta e_A \over e_A})^2,
\eeq
where we have used the fact that the vacuum expectation value of $v_A \sim \sqrt{|e_A|}$.

The $D4$ brane wrapping the dual two-cycle which causes this jump in flux has a tension, 
\beq
\label{tensbu}
T_{brane}\propto |\delta e_A v_A|. 
\eeq
Inserting the correct proportionality factors and converting to Einstein frame as in the previous subsection   now gives, 
\beq
\label{tensebu2}
T_{brane}\sim \sqrt{|V_0|} \delta |{\delta e_A \over e_A}|.
\eeq
Thus comparing eq.(\ref{cbutension}), eq.(\ref{tensebu2}), gives the condition, 
\beq
\label{condbu}
|{\delta e_A \over e_A}|\ll 1,
\eeq
which must be met for the thin wall approximation to hold. 

We now turn to various different cases. In the discussion in \S4.5-\S4.7 that follows we set $m_0>0$ for simplicity.
The case $m_0<0$ can be obtained by changing the sign of all fluxes. 
 
\subsection{Type 2) to Type 2) Decays}

The vacuum energy is given in eq.(\ref{gndse2}).  In this case, $e_A, e_i>0$. 
The change in  blow up fluxes $\delta e_A$ contributes to the difference in energy density $\epsilon$ and thus to $T_u$,
\beq
\label{ctu}
\delta T_U=-\sqrt{3\over 2 \beta } \sqrt{|V_0|}\sqrt{e_A^3\over e_1e_2e_3}\sum_A {\delta e_A\over e_A}.
\eeq
Using eq(\ref{potmin}),eq(\ref{defvol}), eq(\ref{valvi}) and the fact that for Type 2 vacuum $v_A=- \sqrt{10 e_A \over 3 \beta |m_0|}$, 
 we get
\begin{equation}\label{TUtype22}
 T_U = {1 \over \sqrt{2}} {e^{2 D} \over \sqrt{vol} } (\sum_i \delta e_i v_i+  \sum_A \delta e_A v_A).
\end{equation}
In obtaining this formula we had also added the contribution due to the change in the $e_i$ flux which was obtained in eq.(\ref{TUuntw})
 above. 

Since $\Delta W =  \delta e_i v_i + \delta e_A v_A$, and the Kahler potential eq(\ref{kpkm2}) can be approximated  as eq(\ref{kpmod}) to
leading order in $\delta$, we get
\begin{equation}\label{TLtype2}
 T_L = 2 e^{K\over 2} |\Delta W| = {e^{2 D} \over \sqrt{2 vol} } |\sum_i \delta e_i v_i+ \sum_A \delta e_A v_A|.
\end{equation}
For the decay to occur the rhs of eq.(\ref{TUtype22}) must be positive, thus we see  that again  $T_U = T_L$. 
Therefore the decays can again be   at most marginal.

In the calculation above the effects due to the jump in the non-blow up fluxes were calculated as in the previous subsection and 
thus are correct only to leading order in $\delta$. Thus we are assuming that 
\beq
\label{ass2t2}
|\delta e_i v_i| \delta \ll |\delta e_A v_A|.
\eeq
Using the relation that $|{v^A\over v^i}|\sim |{e_i\over e_A}| \delta$ this gives, 
\beq
\label{ass3}
|{\delta e_i \over e_i}| \ll |{\delta e_A \over e_A}|.
\eeq

In fact the Type 2) vacua are stable with at best marginal decays upto a high order of approximation.
This is due to their being related (after a change in the sign of all fluxes) with supersymmetric vacua, as will be explained in section \S4.8.

\subsection{Type 3) to Type 3) Decays}
Here we consider the analogous decays for Type 3) vacua. In this case $e_A<0$, $e_i>0$.  
The ground state energy is given in eq. (\ref{gndse3}). Including a contribution due to the change in the $e_i$ flux gives, 
\begin{equation}\label{TUtype3}
 T_U =   {1 \over \sqrt{2}} {e^{2 D} \over \sqrt{vol} } (\sum_i\delta e_i v_i+  5 \sum_A \delta e_A v_A).
\end{equation}
Note that the contribution proportional to $\delta e_A$ on the rhs comes with a coefficient $5$.
 $T_L$ continues to be  given by eq(\ref{TLtype2}). Therefore now there can be situations where $T_U>T_L$. 

As an example consider the case where $\delta e_i$ vanishes, and one of the $\delta e_A\ne 0$. 
 For the energy difference to be positive, $\epsilon>0$, which means  $\delta e_A<0$, since  $v_A <0$. 
As we will argue below, in this case there is a susy cycle which saturates the lower bound $T=T_L$. 
Thus $T<T_U$ and the decay will proceed.  

The argument establishing that there is a susy cycle with $\delta e_A\ne 0, \delta e_i=0$ is as follows. 
Blowing up the orbifold slightly  gives rise to a $P^2$ at every fixed point. 
There is \footnote{We thank S. Kachru for  suggesting this possibility and I. Biswas and N. Nitsure for extensive discussions.}
 a $P^1\subset P^2$ . 
It is easy to see that this $P^1$ is a 
 holomorphic cycle and  is non-trivial in homology \footnote{In the coordinates used in \cite{GS}, eq.(3.1),
 this cycle is given by setting $z_2=w=0$, so it is clearly holomorphic. 
To include  the point at infinity, $z_1\rightarrow \infty$ a second patch is needed. The 
Kahler form integrates to a non-zero value on this cycle so it is clearly non-trivial in homology.}.
 Its size, $a$,  is proportional to $v_A$ the blow-up modulus. 
Now  being holomorphic $a$ must be proportional to the resulting jump in the superpotential.  This can only happen if the $\delta e_i$ coefficients 
vanish for this cycle, since $v_i\gg |v_A|$ \footnote{Ideally we should have  calculated the intersection numbers of this cycle with 
the $P^2$ divisor and the other four-cycles from first principles 
 and shown that these are of the required form. We will not attempt this  here.}.

Let us estimate the decay rate which results in this case from changing $\delta e_A$.
 Since the tension
\beq
\label{valtfeb}
T=T_L \sim T_U={\epsilon\over \sqrt{3 |V_0|}}
\eeq we  see that
${\epsilon \over T}\sim \sqrt{|V_0|}$, and so from eq.(\ref{exprrho})  the size of the bubble is
\beq
\label{estfeb}
\bar{\rho}\sim {1\over \sqrt{|V_0|}}.
\eeq
The bounce action is $B\sim B_{wall}\sim \bar{\rho}^3 T$. Substituting for $T, \bar{\rho}$ from eq.(\ref{valtfeb}), eq.(\ref{estfeb}), we
then see that the bounce action is
\beq
\label{estten}
B \sim  {\epsilon M_{Pl}^4 \over  V_0^2}
\eeq
where we have reinstated the dependence on the four dimensional Planck scale $M_{Pl}$ on dimensional grounds.

The rate of decay goes like $\Gamma \sim  e^{-B}$, so the fastest decays are those with the smallest jumps in flux. 
Working out the resulting discharge of a particular vacuum due to all the competing decays is a fascinating question 
that we leave for the future \footnote{For this we also need to take into account the fact that inside the bubble is a negatively 
curved FRW universe which ends in a big crunch.}. 

\subsection{Type 3) to Type 2) Decays}

Here we discuss the decays of Type 3) to Type 2) vacua. 
The former have $e_A<0$ while the latter have $e_A>0$. 
Thus the $e_A$ flux must reverse sign in such a decay and we see that   condition eq.(\ref{condbu}), which we had imposed so that the 
variation of the $v_A$ moduli does not contribute significantly to the domain wall tension,  cannot be met. 
This situation is analogous to  \S4.1 where we dealt with decays from non-susy to susy vacua. 
The expectation value of the $v_A$ moduli depend actually on the absolute value of $e_A$. 
So to meet  the thin wall approximation we can   now adjust 
$|e_A|$ so that the variation in $v_A$ is small and therefore its contribution to the domain wall tension can be neglected. 
To illustrate this we infact adjust $|e_A|$ so that this variation vanishes. 

Using $v_A=-\sqrt{10 \over 3} \sqrt{|e_A| \over  \beta |m_0|}$ for Type 2) and $v_A=-\sqrt{10\over 21} \sqrt{|e_A| \over  \beta |m_0|}$ for Type 3,
we learn that for $v_A$ to be the same,
\begin{equation}\label{eA}
 e_A \ _{\text{type 3}} = - 7   e_A \ _{\text{type 2}}.
\end{equation}
Using this, we can calculate the difference in energy 
\begin{equation}
\epsilon = V_{\text{type 3}}- V_{\text{type 2}}= {3 \over 2} \sum {\delta e_i \over e_i}  |V_0|  + c |V_0| 
\sqrt{\kappa \over\beta  e_1 e_2 e_3} \sum_A |e_A \ _{\text{type 2}}|^{3 \over 2}.
\end{equation}
where $c=\sqrt{2} +7 \sqrt{50}$.
Note that for the decay to be possible $\epsilon>0$. 
 Using eq(\ref{rel1})  $T_U$ can be calculated to be
\begin{equation}\label{TUtype23}
T_U =  \left(   \delta e_i v_i  +24 \sum_A  e_A \ _{\text{type 2}} \ |v_A| \right) {e^{2 D} \over  \sqrt{2 vol}}.
\end{equation}
Note that for $\epsilon>0$ the term in the brackets in the above equation is greater than zero.

$T_L$ can be calculated to be,
\begin{equation}
 T_L = |\delta e_i v_i + \sum_A \delta e_A v_A | {e^{2D} \over \sqrt{2 vol}}.
\end{equation}
Now $v_A<0$ and  from eq(\ref{eA}) we know that $\delta e_A = -8 e_A\ _{\text{type 2}}$,  therefore
\begin{equation}
\label{se23jan}
 T_L = |\delta e_i  v_i + 8 \sum_A e_A\ _{\text{type 2}} |v_A| )| {e^{2D} \over \sqrt{2 vol}}.
\end{equation}
It is now clear that as long as $\delta e_i v_i>0$, $T_U>T_L$. A decay will be allowed if $T<T_U$. 
Like in the  Type 3) -Type 3) case,  concrete examples can be given where this is true. 
 For example in  the case where $\delta e_i=0$ the $D4$ brane tension  saturates the lower bound with $T=T_L$,
since it is a holomorphic cycle, leading to an allowed decay.

We can also ask  about the possibility of Type 2) vacua decaying to Type 3). Running the above argument again the coefficient
$24$ in the second term on the rhs of eq.(\ref{TUtype23}) and $8$ in the second term of eq.(\ref{se23jan}) both reverse sign making both these terms negative. Since $\epsilon>0$ for the decay to happen, we find  that  
$T_U<T_L$. This shows that  such decays are disallowed.  
 
We have adjusted the fluxes so that the $v_A$ moduli have the same value in the two vacua, thereby ensuring that the moduli contribution to the domain wall tension is small.  Our conclusions will remain  unchanged  if the fluxes took different values, allowing for a variation in $v_A$, as long as one
 stays in the thin wall approximation.

\subsection{Supersymmetric Partners in the Landscape and Marginality}\label{sec2.3}

We have seen above that Type 2) vacua are stable and can at most decay marginally, within our approximations. 
We will now see that this stability is quite general  and 
 can be understood by relating these vacua to supersymmetric ones. 

We had seen in \S2.1, eq.(\ref{potential}), that when only $e_i$ fluxes are excited    the potential energy is invariant under 
a change in sign of the four-form fluxes, $e_i\rightarrow -e_i$, as long as the axions,
$b_a$ all vanish. In fact this is more generally true and  follows directly from the IIA supergravity action where  the $F_4$ dependence arises in the  term,
\beq
\label{iiaterm}
S_{IIA}=-{1\over 2}\int d^{10}x\sqrt{-g}  |\tilde F_4|^2 + \cdots
\eeq
with,
\beq
\label{deff4t}
\tilde{F}_4=F_4-F_2\wedge B_2-{m_0\over 2} B_2\wedge B_2,
\eeq
As long as $B_2$ vanishes\footnote{More correctly we mean the axions which arise from $B_2$ should vanish.}, 
taking  
\beq
\label{symmrev}
F_4\rightarrow -F_4,
\eeq
 gives the same action.

In contrast the conditions for supersymmetry {\it do}  care about the sign of the fluxes, as we have discussed extensively above. 
Now in the Type 2) vacua all the four- form flux has a sign opposite to that required by  supersymmetry. 
This means that starting with a vacuum of Type 2) we can construct a susy  vacuum with the same energy by reversing all the $F_4$ fluxes. 
This susy vacuum  will also have the same expectation values for the Kahler moduli and the dilaton.

Now consider a possible decay of  a Type 2)  vacuum to  a susy vacuum of this type. By reversing the sign of all the fluxes we can
relate this to the decay of a susy vacuum to a non-susy Type 2) vacuum. The vacuum energies of the initial and final vacua in the 
first decay and its partner decay are the same. The domain wall in the second case carries charges opposite to the 
first one. If the first decay is mediated by 
a $D4$ brane wrapping some  cycle, the partner is mediated by the anti D4 brane wrapping the same cycle. 
Thus the two domain walls must also have the same tension. 
It then follows that the first decay of the non-susy vacuum can be allowed iff the partner susy vacuum can decay. 
But on general grounds one expects the susy vacua to be stable. 
We therefore  conclude that the Type 2) non-susy vacuum we started with also cannot decay. 

It is clear that a similar argument would also work if instead of considering the decay of the Type 2) vacuum to a
 susy vacuum we   considered its  decay to another Type 2)
or a Type 3) vacuum. In both of these cases  the axions are not turned on. By reversing all the four- form fluxes we can relate this to the 
decay of the susy vacuum to a susy  vacuum in the first case, or the decay of a susy vacuum to the partner of a Type 3) vacuum in the second case. 
Both should not occur, given the stability of susy vacua.  The partner for the Type 3) case is a vacuum with $e_i<0,
e_A>0$. This is case 4) in the classification of \S2. Here 
 the $b_A$ axions are tachyonic and lie below the BF bound, but this does not invalidate the argument above. 

How general is this argument which ensures the stability of the Type 2) vacua by relating it to partner susy vacua? 
Our discussion above is based on the thin wall approximation in supergravity. And holds if the true and false vacuum have vanishing values
for the axions. In the thin wall approximation only the $D4$ brane contribution to the domain wall tension is important, and this is clearly 
the same in the non-susy vacuum decay and its partner.  
  Going beyond, one can argue that the domain wall tension continues to be equal in the two cases if the moduli 
contribution is included in the tension, as long as the axions are not activated in the domain wall. This follows from the fact that the 
potential energy and Kinetic energy terms all respect the flux reversal symmetry in the absence of axions. 
  Since the axions vanish in both the true and false vacuum there is no reason as such for them to get activated, but  for thick enough
walls where the moduli under go big excursions this could happen  anyways as a way of reducing the tension. If so,  eventually for a thick enough 
wall the argument would break down.   

 Even for decays which are well described by the thin wall approximation subleading correction are important in the marginal case. 
We had found above that decays of Type 2) to Type 2) vacua are marginal if the $D4$ brane wraps a susy cycle. This result is easy 
to understand in light of the above discussion, since   the partner susy decay would be now mediated by a BPS domain wall. 
However in the non-susy  Type 2) decay case, one expects that  the marginal nature is only approximate and eventually 
corrections lead to the   decay being either allowed or disallowed. 
The   corrections responsible for this might  arise as corrections to the thin wall approximation itself, as we have mentioned above, 
or they might  require going beyond the sugra approximation and including $\alpha'$ and $g_s$ corrections. 
We leave an exploration of such questions for the future. 

Finally the argument above applies only for decays of the Type 2) vacua to others where the axions are not turned on.
All the stable vacua we have explored in this paper are of this type, but there could be other vacua where the $b_A$ axions have non-zero 
expectation values. The argument above says nothing about the possible decays of Type 2) vacua in such cases and this 
would have to be examined on a case by case basis.

\subsection{More on Supersymmetric Domain Walls}
Let us end this section with some more comments on susy domain walls. We had mentioned in the discussion around eq.(\ref{ineq}) that 
$D4$ branes which wrap holomorphic or antiholomorphic cycles preserve susy. 
More accurately if we take Type IIA  on 
the Calabi-Yau manifold without flux the $D4$ brane wrapping such a cycle will preserve half the supersymmetries, i.e., ${\cal N}=1$.
If we now turn on flux to preserve ${\cal N}=1$ susy then only one of the two cases, either the holomorphic or antiholomorphic cycle,
preserves the surviving ${\cal N}=1$ susy \cite{Aharony1}. That only one of the two cases could preserve susy at best is easy to understand from the 
requirement of force balance.  The antiholomorphic case can be thought of as the anti D4 wrapping the same cycle. 
If the attractive gravitational force cancells the RR repulsion  for the brane it will not cancell for the anti-brane and 
vice-versa.

It is easy to see that a susy brane  leads to a marginal decay. In this case the tension is given by $T_L$ and the energy difference,
$\epsilon=-3e^K\Delta|W|^2$. It is then easy to see that the condition for marginality,
\beq
\label{condmarginality}
{\epsilon \over 3 T}-{ T \over 4}=\sqrt{ |U| \over 3},
\eeq
is met, where $U$ is the cosmological constant.
The tension is given by $T_L$ in the probe approximation. 
Going beyond would require including  changes in the moduli which arise because the brane causes the flux to jump.
One expects the susy branes to continue to be marginal even then. Susy domain walls  where moduli fields vary have been discussed in
\cite{Weinberg}, \cite{CveticRey}, \cite{Cvetic}, \cite{Kallosh3}, where it was shown that the walls are indeed marginal
\footnote{
This is true only  when the superpotential does not vanish in between the two vacua, otherwise the wall tension is too big. In our case,
starting with the probe approximation and including corrections, the superpotential will not vanish. However
for larger changes
the resulting analysis might be more involved.}.  In this analysis the fluxes (which are parameters in the superpotential) were held fixed.
One could try to include the changes of flux in the analysis of these authors as well, but we leave this for the future.

Finally, in practice 
 given the charges carried by the domain wall it is not always easy to decide whether a corresponding
 supersymmetric cycle exists.
As a special case we can consider the orbifold theory and linear branes,  for which the $z^i$ coordinates are
linear functions of $\sigma, \bar{\sigma}$. Even in this simple case, the existence of  a supersymmetric cycle
 translates into a fairly intricate number theoretic constraint on the wrapping numbers of the $D4$ brane, as discussed in \cite{Aharony1}.
Things simplify if the integers $\delta n_i$ are large, $|\delta n_i| \gg 1$.
Now, upto fractional corrections, which are of order $1/\sqrt{|n|}$, we can approximate,
 $\delta n_i \simeq\pm m_i^2$, to be a perfect square. The only obstruction to having a susy  brane then arises due to the signs of the $\delta n_i$.
If the $\delta n_i$'s all have the same sign then a susy cycle exists, else it does not exist. 

\section{More General Decays}
In this section we consider domain walls which carry more general charges. 

The general vacuum with all fluxes turned on was discussed in \S2.4.  The ground state energy for different vacua  can be calculated
by replacing $e_a$ in formulae obtained in the case with $m_a=0$ , with eq.(\ref{defhate}). 

Our discussion of domain walls will follow that in \S4 above. Given a domain wall with some charges, 
the change in the superpotential provides a lower bound on its tension.
Below we will then calculate this lower bound, $T_L$ and compare with an upper bound $T_U$ defined in eq.(\ref{TU}). 

We calculate $T_L$ by keeping the moduli which appear in the superpotential to be fixed.
We will come back to justifying this probe approximation below when we also discuss the validity of the thin wall approximation. 
For the superpotential, eq.(\ref{fullsupjan}), 
the change due to a domain wall carrying charges, $(\delta e_0, \delta e_a, \delta m_a)$ is, 
\beq
\label{chngwjan2}
\Delta W=\delta e_0 + \delta e_at_a+\kappa_{abc}\delta m_a t_bt_c.
\eeq
It is useful to express this in terms of the real and imaginary parts of $t_a=b_a+iv^a$, and in terms of $\hat{e}_a$, 
\beq
\label{defdelhat}
\delta \tilde{e}_a=\delta e_a + m_0 \kappa_{abc}\delta m_b b_c.
\eeq
This gives,
\beq
\label{delw2}
\Delta W=\delta e_0+\delta e_a b_a + {{\kappa_{abc}}\over 2} \delta m_a (b_bb_c - v_bv_c) + i \delta \hat{e}_a v_a.
\eeq

From here using eq.(\ref{lb2jan}) we get,
\beq
\label{tlgen}
T_L=2e^{K/2}|\Delta W|={e^{2D}\over \sqrt{2 vol}}\sqrt{(\delta\hat{e}_a v_a)^2+([\delta e_0+\delta e_a b_a+{\kappa_{abc} \over 2}
\delta m_a b_bb_c] - {\kappa_{abc} \over 2} \delta m_a v_b v_c )^2}.
\eeq
 The axion fields $b_a$  which appear above   have a vacuum expectation  value,
\beq
\label{vevba}
b_a={m_a\over m_0}.
\eeq
 
It is worth pausing to discuss the physics behind this expression. 
The $6$-brane component of the domain wall gives rise to an induced $4$ brane component because the axions $b_a$ are now-non-zero.
This is responsible for the shift in $\delta e_a$, eq.(\ref{defdelhat}). The second term within the square root arises from a $D2$ brane
and a $D6$ brane component. The $D2$ brane component included a contribution due to induced  $D2$ brane charge which arises from the 
$D4$ brane and $D6$ brane components in the presence of the axions. These together with $e_0$ account for the  term 
 within the square brackets.

Let us now compare this with $T_U$. 

For a Type 2) - Type 2) decay this is given by replacing $(\delta e_i, \delta e_a)$ in eq.(\ref{TUtype22}) by their hatted counterparts 
giving,
\beq
\label{genTU22}
T_U={e^{2D}\over \sqrt{2 vol}} |\delta \hat{e}_a v_a|.
\eeq
Thus we see that  for this case now $T_L\ge T_U$ and the decay is at best marginal. 
The marginal case arises when the second term within the square root in eq.(\ref{tlgen})  vanishes. For this to happen
the sum total of the $D2$ brane charge and 
$D6$ brane  must vanish. In addition  the tension must  equal  the lower bound,
this would require the brane configuration to be supersymmetric.

For Type 3) -Type 3) we obtain $T_U$ by replacing the $\delta e_a$  by $\delta \hat{e}_a$ in eq.(\ref{TUtype3}).
This gives, 
\beq
\label{genTU33}
T_U={e^{2D}\over \sqrt{2 vol}}|\delta \hat{e}_i v_i + 5 \delta \hat{e}_A v_A|.
\eeq
Now we see that $T_U$ can be greater than $T_L$. For example this can happen if the sum of the D2 and D6 brane charges vanish and the $\delta e_i$ and $\delta e_A$ fluxes have opposite sign \footnote{Remember that $v_i, v_A$ have opposite signs, so when this happens the two contributions actually add.}.
In such cases if the tension is equal to $T_L$ or close to it the decay will occur. 

Similarly one finds  that decays of Type 3) to Type 2) can indeed occur. 
And also one finds that decays of Type 2) to Type 3) cannot occur because $T_L>T_U$.
We skip some of the details here. 

Let us end with three comments. 
First, in this section we have not discussed the constraints imposed by the thin wall approximation. This requires that the 
contribution the moduli make  to the domain wall tension is smaller than the 
$D$ brane contribution. It is straightforward  to evaluate 
 the moduli contribution and then impose this constraint; we will spare the patient reader the
tedious details. 

Second, one of the conclusions that follows from our analysis above is that  Type 2) vacua continue to be stable even after decays 
involving the most general kind of brane are considered. We had argued in \S4 that there was a pairing symmetry 
which related the Type 2) vacua to susy vacua and this explained 
their stability. However this symmetry required that the $b_A$ axions are not turned on. 
Now, with the most general kind of interpolating brane,  $m_a$ will in general undergo a  change so that the $b_a$ fields 
will become non-zero even if they vanish to begin with. 

It turns out that while there is no exact symmetry which relates Type 2) vacua to susy ones  in general,
there is  an approximate symmetry of this type to leading order in the change in fluxes, as is discussed in Appendix D.
 The calculations above have been carried out 
only at leading order.   For example we have calculated the change in vacuum energy density $\epsilon$ and thus $T_U$ to leading order in 
$\delta e_a$ and $\delta m_a$. Thus this approximate symmetry provides an explanation for the  stability of the Type 2) vacua in our approximations. 
Subleading corrections beyond first order would not be important for decays which are ruled out at leading order. 
However they are important in the marginal case. The fact that  the symmetry is only an approximate one for the general case  suggests
 that Type 2) vacua  might well decay by nucleating a brane which includes  a $D6$ brane component once such subleading effects are incorporated. 
We leave a detailed investigation of this question  for the future. 

Finally, one can also consider $D8$ and $NS5$ brane mediated decays. These lie outside the thin wall approximation. This too is discussed in Appendix D.

\section{Discussion}

\smallskip
\noindent
$\bullet$
We have constructed two explicit classes of non-susy AdS vacua in this paper, denoted as Type 2) and Type 3).
Both are perturbatively stable. 
We have found that several possible decays of these vacua to other susy and non-susy vacua with lower energy are disallowed since the tension of the interpolating domain
wall is much too big. The underlying reason for this is the geometric nature of AdS space where volume and surface area grow at the same rate for a large bubble. 

The Type 3) vacua do have   allowed decays   to  some other Type 3) and Type 2) vacua. 
The   Type 2) vacua  were found to be   stable in our approximations, although some decays are only marginally disallowed. 
It is important to go beyond our approximations to decide what happens in these marginal cases.
By changing the sign of all the four-form fluxes the Type 2) vacua are turned into susy vacua with the same energy.
We argued, within our approximations, that the stability of the susy vacua then ensures the stability of their Type 2) partners. 
This  protection mechanism   might well be more robust and perhaps extends even  beyond leading order, but one expects it to  eventually fail, 
 tipping
 the marginal decays one way or another. 
We leave an analysis of this for the future. 

\smallskip
\noindent
$\bullet$
Our analysis was carried out  by considering a specific model of IIA theory on the blown-up $T^6/(Z_3\times Z_3)$, after including the effects of 
flux and a further orientifolding. However some of our conclusions are more general and apply to  IIA on  any Calabi-Yau manifold with fluxes. 
 E.g., Type 2) vacua, obtained by flipping the sign of all the four-form fluxes exist as  extremum of the potential in general, since   in 
the absence of  axions coming from the $B_2$ field,  IIA sugra   will continue to have the symmetry, 
eq.(\ref{symmrev}). 
However their perturbative stability is not guaranteed in general, since some of the axions could lie below the BF bound in these vacua.

\smallskip
\noindent
$\bullet$
A small bulk rate of decay leads to a diverging decay rate in the boundary as we had discussed in the introduction. 
What is the dual description of this in the boundary CFT? 
In the bulk the  divergence  arises after  integrating over all radial locations of the instanton, due to the diverging bulk volume.
It is tempting to speculate that in the boundary there is a corresponding one-parameter family of instantons, 
parametrised by their size. And summing over the different sizes   then gives rise to this divergence, 
which arises in the CFT due to  instantons of very small size \footnote{We are grateful to G. Mandal,  S. Kachru and 
 S. Minwalla for a discussion of these issues.}.

It might seem that the the divergence mentioned above can 
  be controlled by introducing a cut-off  at a large and finite radial location in the bulk. 
Conformal invariance would  not be  exact now but would be  an approximate symmetry in the deep IR. 
However a more detailed analysis is needed, depending on the kind of instability one is dealing with,
before one can be sure.   It could be that the  detailed nature of the 
boundary conditions at the cut-off play a significant role even in the IR \footnote{
This is more of a worry, in our minds, for the kind of decays discussed in this paper 
nucleated by a $D$ brane rather than decays in non-susy orbifolds 
\cite{HOP} nucleated by an instanton analogous to the one responsible for the decay of the KK vacuum \cite{Witten2}. 
In the $D$-brane case the RR repulsion dominates over gravitational attraction and results in    
a runaway  $-\phi^6$ potential arising in its world volume action.  
This could potentially cause an instability in the quantum theory whose cure depends delicately 
on the correct boundary conditions.}. We leave a detailed understanding of this divergence in the boundary theory and related issues about controlling it also for the future.

\section{Acknowledgements:}
 We thank I. Biswas, D. Harlow, S. Kachru, G. Mandal, S. Minwalla, N. Nitsure, A. Sen, S. Shenker and  E. Silverstein for discussion. 
 
S.T.  acknowledges support from Stanford University and Slac during his sabbatical visit in 2008-2009 when this work was started. 

We acknowledge funding support from the DAE, Government of India.  Most of all we thank the people of India for enthusiastically supporting 
research in string theory.

\section{Appendix A : Blow up modes }
In this appendix we give the details of the  calculations involved in \S2.3. We give the potential after including the blow up modes and discuss its minimization.
We work in the regime where $|{e_A \over e_i}|\ll 1$. The potential can be then 
 expanded in the small parameter $ {|e_A|^{3/2} \over |e_i|^{3/2}}$.

The superpotential W and Kahler potential K are given as in eq(\ref{kpkm2}) and eq(\ref{superpot2}). We have already found in eq(\ref{massmatrix}) that with $e_A=0$, 
the axions of the untwisted sector which vanish  at the extrema lie above the BF bound. Now their masses will receive corrections of order 
$ {(|e_A|/ |e_i|)}^{3/2}$ but these are small and so their masses  will continue to lie above the B.F bound. 

The various Kahler derivatives to the leading order are as follows

\begin{eqnarray}\label{kahlerd}
\nonumber
K_{t_i \bar{t}_i} = {1 \over 4 v_i^2} \left( 1 - {\beta \sum v_A^3 \over 3 vol }\right).   &\hspace{10mm}&K_{t_i \bar{t}_i}  = - {\beta \sum v_A^3 \over 24 vol} {1 \over v_i v_b}. \\
K_{t_A \bar{t}_B}=- {\beta v_A \over 4 vol}  \delta_{AB}.&  &K_{t_A \bar{t}_i} = {\beta v_A^2 \over 8 vol v_i }.
\end{eqnarray}
Their inverses are
\begin{eqnarray}\label{kahlerdi}
\nonumber
K^{t_i \bar{t}_i} =  4 v_i^2 ( 1 +  {\beta \sum v_A^3 \over 12 vol }).   &\hspace{10mm}& K^{t_i \bar{t}_i}  = -  4 v_i v_b ( {\beta \sum v_A^3 \over 12 vol}) .\\
K^{t_A \bar{t}_B}=- {4 vol  \over \beta v_A} \delta_{AB}.&& K^{t_A \bar{t}_i} = 2 v_i v_A.
\end{eqnarray}
The derivatives of superpotential as
\begin{eqnarray}\label{covderWK}
\nonumber
D_{t_i} W &=& \partial_{t_i}W + K_{t_i}W=  \left( e_i + {m_0 vol \over v_i} - {W\over 2 i v_i} (  1 - {\beta \sum v_A^3 \over 6 vol } )  \right) . \\
D_{t_A}W &=& \partial_{t_A}W + K_{t_A}W = e_A+{m_0 \beta v_A^2 \over 2} ( 1 - i {b_A \over v_A})^2 - {\beta v_A^2 \over 4 i vol} W.
\end{eqnarray}
Using these we can calculate the potential from
\begin{equation}
\label{potform}
V = e^K \left( K^{t_a \bar{t}_b} D_{t_a} W D_{\bar{t}_b} \bar{W} - 3 |W|^2\right).
\end{equation}
We keep terms in the potential to leading order in ${|e_A| \over |e_i|}^{3 \over 2}$. 
Our analysis will show that $v_A , b_A \sim \sqrt{|e_A|}$. Keeping all terms to this order then gives, 
\begin{eqnarray}\label{potvAbA}
\nonumber
V &= & {e^{4 \phi} \over 2 vol^3 }  \sum_i e_i^2 v_i^2 + {p^2 \over 4} { e^{2 \phi} \over  vol^2}+ \sqrt{2} m_0 p {e^{3 \phi} \over vol^{3 \over 2}} + {m_0^2 \over 2} {e^{4 \phi} \over vol} \\
\nonumber
&&- \sum_A {e_A^2 \over \beta v_A vol } {e^{4 \phi} \over  2 vol} + \sum_A ({e_A v_A \over vol} ) \sum_i(e_i v_i) {e^{4 \phi} \over 2 vol^3}  \\
\nonumber
&&- {\beta \sum v_A^3 \over vol} \left( {5 e^{4 \phi} \over 24 vol^3}  \sum_i e_i^2 v_i^2  + {e^{4 \phi} \over 12 vol^3} \sum_{i\ne j} e_i v_i e_j v_j \right. \\
\nonumber
&& \left. -{p^2 e^{2 \phi} \over 12 vol^2} + \sqrt{2} {e^{3\phi} m_0 p \over vol^{3 \over 2} } + {m_0^2 e^{4 \phi} \over 12 vol} \right) \\
\nonumber
&& + \sum_A {b_A^2 \over vol}  \left\lbrace {e^{4 \phi} \over 2 vol} {m_0 e_A \over v_A} - {3 e^{4 \phi} \over 8 vol^2} \sum_i (e_i v_i)m_0 \beta v_A + {3 m_0 p  e^{3 \phi} \over 4 \sqrt{2} vol^{5 \over 2}} \beta v_A \right. \\
&& \left.- { e^{4 \phi} \over 8 vol} m_0^2 \beta v_A \right\rbrace.
\end{eqnarray}
The minima of the zeroth order potential were given in eq(\ref{valvi}),eq(\ref{vald}) and eq(\ref{valphi})\footnote{These may receive ${|e_A| \over e_i}^{3 \over 2}$ corrections, but they are immaterial for the potential to the leading order.}. Plugging these into the eq(\ref{potvAbA}), we get
\begin{eqnarray}\label{Vblowup}
\nonumber
 V &=& V_0 + |V_0| \sqrt{15 \kappa \over 4 \beta} \sum_A \sqrt{|e_A|^3 \over |e_1 e_2 e_3|} \left\lbrace {1 \over 2 x_A} -  { 3 \: x_A \: \text{sign}(e_A) \sum_i \text{sign}(e_i) \over 10}  \right. \\
\nonumber
&&\left. +  {3 x_A^3 \over 200} (1 + 2 \sum_{i \ne j} \text{sign}(e_i e_j) )  + {y_A^2 \over 40} \left[ - {20 \text{sign}(m_0 e_A) \over x_A}  \right. \right. \\
&& \left. \left. + x_A ( 29 + 9 \sum_i \text{sign}(m_0 e_i)) \right]\right\rbrace,
\end{eqnarray}
where $x_A = -v_A \sqrt{\beta |m_0| \over |e_A|}$ and $y_A = b_A \sqrt{\beta |m_0| \over |e_A|}$.
Thus the minimization of the blow up modes will depend upon the signs of various fluxes turned on. 

Table(\ref{cases}) lists the different cases which arise from  the different choices for signs of fluxes. 
For convenience  we take $m_0>0$. For the other choice, $m_0<0$, results can be obtained by changing the sign of all the $e_a$ fluxes. 
In the Table the nature of the extremum, if present, along the $x_A$ direction is discussed in the second last column. 
The potential obtained for the $y_A$ field after setting $x_A$ to its extremum value is in the last column. 
\begin{table}\begin{center}
\begin{tabular}{|c||c|c|c|c|}
\hline
Case &$e_1,e_2,e_3$ & $e_A$ & Extrema of $x_A$,nature & Potential  for $y_A$\\
\hline
1.&$-,-,-$&$-$&$\sqrt{10 \over 3}$, minima& ${1 \over \sqrt{2}} \sqrt{\kappa \over  \beta}\sqrt{|e_A|^3 \over |e_1 e_2 e_3|} |V_0| y_A^2$ \\
\hline
2.&$+,+,+$&$+$&$\sqrt{10 \over 3}$, minima&${25 \over4 \sqrt{2}}\sqrt{\kappa \over  \beta} \sqrt{|e_A|^3 \over |e_1 e_2 e_3|}|V_0| y_A^2$\\
\hline
3.&$+,+,+$&$-$&$\sqrt{10 \over 21}$, minima&${7 \sqrt{7} \over 4 \sqrt{2}}\sqrt{\kappa \over  \beta}\sqrt{|e_A|^3 \over |e_1 e_2 e_3|} |V_0| y_A^2$ \\
\hline
4.&$-,-,-$ &$+$&$\sqrt{10 \over 21}$, minima& $ - {5  \over \sqrt{14}} \sqrt{\kappa \over \beta}\sqrt{|e_A|^3 \over |e_1 e_2 e_3|} |V_0| y_A^2$\\
\hline
5.&$+,+,-$&$+$&No extrema & \\
\hline
6.&$-,-,+$&$-$&No extrema & \\
\hline
7.&$+,+,-$ &$-$&$\sqrt{10 \over 3}$, Inflection&$  { 11 \over 2 \sqrt{2} }\sqrt{\kappa \over  \beta}\sqrt{|e_A|^3 \over |e_1 e_2 e_3|} |V_0| y_A^2$ \\
\hline
8.&$-,-,+$&$+$&$\sqrt{10 \over 3}$, Inflection& $ {7\over 4 \sqrt{2}} \sqrt{\kappa \over  \beta}\sqrt{|e_A|^3 \over |e_1 e_2 e_3|} |V_0| y_A^2$\\
\hline
\end{tabular}
\caption{\textbf{Different choices of signs with $m_0>0$. The $\pm$ denotes the sign of the corresponding flux.}}
\label{cases}\end{center}
\end{table}\\

\smallskip
\noindent
$\bullet$
Case 1 is the susy solution found by \cite{DGKT}, who also discuss some aspects of the non-susy solutions. 

All the other cases below break supersymmetry. 

\smallskip
\noindent
$\bullet$
Case 2) and Case 3) are perturbatively stable vacua. 

\smallskip
\noindent
$\bullet$
Case 4 : The blow up mode has a tachyon.  To see whether it is acceptable we compare its mass with the BF bound. Note that in this case the action for $y_A$ is
\begin{eqnarray}
 \label{actionax}
\nonumber
S &=& \int {dx}^4 \left(   K_{t_A \bar{t}_A} \partial_\mu t^A \partial^\mu \bar{t}^A- {5  \over \sqrt{14}} \sqrt{\kappa \over \beta}\sqrt{|e_A|^3 \over |e_1 e_2 e_3|} |V_0| y_A^2  \right)\\
&=&  \sqrt{\kappa \over \beta}\sqrt{|e_A|^3 \over |e_1 e_2 e_3|} \int {dx}^4  \left( {3\sqrt{2} \over 20 \sqrt{7}}  \partial_\mu {y_A} \partial^\mu y_A - {5 \over \sqrt{14}} |V_0| y_A^2 \right).
\end{eqnarray}
Therefore mass of the blow up axion in this case is
\begin{equation}
\label{relpp}
 M_{\text{axion}} = -{50 \over 3} |V_0|. 
\end{equation}
Since the BF bound is $M_{BF} = - {3 \over 4}|V_0|$ , the BF bound is violated.

\smallskip
\noindent
$\bullet$
Case 5 and 6 have no extrema at all (near the orbifold point). 

\smallskip
\noindent
$\bullet$
Case 7. and Case 8. have  extrema for  $x_A$ at a point of inflection (first and second derivatives vanish). Therefore in these cases
the next correction must be calculated to find  if there is a minimum close to the point of inflection. This is determined  by the sign of the 
the next order correction. We  do not carry out this calculation here.

\section{Appendix B: Moduli Tension}


In this section we estimate the contribution the moduli make to the domain wall tension. 
Let $\phi$ be a canonically normalized scalar field
with action,
\beq
\label{actscalarjan}
S_{scalar}=\int d^4x[-{1\over 2} (\partial \phi)^2 -U(\phi)].
\eeq
Consider a domain wall in flat space which is translationally invariant along $x,y$ directions. The scalar field varies along the $z$ direction in the domain wall. 
The tension of the wall receives a contribution from the gradient term and the potential energy term,
\beq
\label{tensiondomain}
T=\int dz [{1\over 2} ({d\phi \over dz})^2 + U(\phi)].
\eeq
The scalar profile is obtained by balancing these two. If  the scalar has a mass $M$, the spatial variation is of order $1/M$. 
For two vacua which are separated by $\Delta \phi$ in field space the resulting tension is then of order, 
\beq
\label{resten}
T\sim M (\Delta \phi)^2. 
\eeq

Strictly speaking in making the estimate, eq.(\ref{resten}),  we are assuming  that the two vacua separated by the domain wall 
are "close by" in field space and  that the mass $M$ for fluctuations about both of them  is
 approximately the same. 
As we discuss in the main text the thin wall approximation within which we work requires that the moduli contribution to the tension is much 
smaller than the D-brane contribution. This requirement   often leads to the condition that  the total change in the scalar,  $\Delta \phi$,
 is sufficiently small, and therefore that  the vacua  are close by.
Even when this is not true and the vacua are not sufficiently close by, the  estimate in eq.(\ref{resten}) for the tension 
is often parametrically a good one.

In our case the   domain walls which arise actually have two components,  a D-brane component which is thin
 and  a component which arises because the moduli vary across them.
In such a situation the potential seen by the moduli  jumps as one crosses the thin $D$ brane, due to the jump in
flux.  Despite this complication, as we will see below, eq.(\ref{resten}) continues to be a good approximation to the tension.

Let us construct the scalar profile for such a domain wall in more detail. 
We take the D-brane, which is thin, to be located at $z=0$ and the two vacua to be close together in field space for simplicity. 
Then the resulting potential can be 
approximated as, 
\begin{eqnarray}
 \label{potquad}
\nonumber
  U(\phi) &=& U_- + M_-^2 \left( \phi - \phi_- \right)^2  \ \ z<0  \\
  U(\phi) &=& U_+ + M_+^2 \left( \phi_+ - \phi \right)^2 \ \ z>0.
\end{eqnarray}
Here $U_{\pm}$ are the values of the  minimum values of potential  to the left and the right  of the wall, 
and $\phi_{\pm}$ are the vacuum expectation values of the scalar field  and $M_{\pm}^2$ are the masses   about these
minima. 
Solving for the scalar with the correct boundary conditions gives 
\begin{eqnarray}
 \nonumber
\phi(z) - \phi_- &=& A e^{M_- z}  \ \  z<0 \\
\phi(z) - \phi_+ &=& - B e^{- M_+ z} \ \  z>0.
\end{eqnarray}
Continuity of $\phi$ and its first derivative at $z=0$ then leads to, 
\begin{eqnarray}
 \nonumber
\phi_- + A &=& \phi_+ - B\\
M_- A &=& M_+ B.
\end{eqnarray}
From these we get that 
\beq
\label{valsbjan}
A={(\phi_+-\phi_)M_+ \over M_+ + M_-} \ \ B={(\phi_+-\phi_)M_- \over M_+ + M_-}.
\eeq
If $\Delta \phi = \phi_+-\phi_-$ is small  we can approximate $M_+\sim M_-$ to leading order and  then it is easy to see that the 
 resulting tension  is  of order 
eq.(\ref{resten}).

\section{Appendix C: More On  Four-Form flux and Normalization of Kahler form}
\subsection{The Jump in the Four Form}

Let us  consider the orbifold limit and take a single $D4$-brane which wraps the first $T^2$, spanned by the $(x_1,y_1)$ directions.
The first $Z_3$ symmetry (denoted by S in \S2)   has fixed points.  We take the $D4$-brane to be located, in the remaining internal directions 
$(x_2,y_2, x_3,y_3)$,  away from these fixed points. The $Z_3\times Z_3$ orbifold action then 
does not result in any self identifications on the world volume of the $D4$-brane. 
Such a $D4$-brane 
 corresponds to winding number $\delta n^1=1$ in our notation. We would like to calculate the  related jump in the $F_4$ flux.

The $D4$-brane is electrically charged with respect to $F_6$ which is governed by  the action 
 \begin{equation}
\label{actd4jan}
 S = - {1 \over 2 \kappa_{10}^2 } \int dx^{10} \sqrt{-g} {F_{6}}_{ABCDEF} F_6^{ABCDEF} + \sqrt{2} \mu_4 \int C_5. 
\end{equation}
The last term arises from the world volume action of the $D4$, and the extra factor of $\sqrt{2}$ in front of this term is because we use the notation of \cite{DGKT} in which  $C_{RR}={C_{RR}^{Polch}\over \sqrt{2}}$, where $C_{RR}^{Polch}$ refers to the conventions of \cite{Polchinski}. 
We will use this action to calculate the jump in $F_6$ and from there the jump in $F_4$. 

The non-compact directions are those of $AdS_4$ with metric 
\beq
\label{metricads}
ds^2=dr^2 {R^2 \over r^2} + {r^2 \over R^2} (-dt^2 + d\mu_1^2 + d\mu_2^2).
\eeq
We work in the orbifold limit and the internal space has  the metric eq.(\ref{metric}).  

The $D4$ brane extends along $(t, \mu_1,\mu_2)$ of the non-compact directions and is located at $r=r_0$. 
It results in a change in the  $(t r \mu_1 \mu_2 x_1 y_1)$ component of the six-form.  
For simplicity we take the $D4$ brane to be smeared out in the remaining internal directions, i.e. the $(x_2,y_2)$ and $(x_3,y_3)$ directions.
Then the six-form sourced by the $D4$ brane will be  independent of all the internal directions. 
Its jump can be calculated by carrying out the integral over the internal space in eq.(\ref{actd4jan}) and working with the resulting $4$ dimensional action. 

This gives, 
\begin{eqnarray}
\label{actd44jan}
S& = & -{1\over 2} ({1\over 9 }) ({\sqrt{3}\over 2})^3 (\gamma_1 \gamma_2\gamma_3) 
\int d^4 x \sqrt{-g_4} F_{6  \ t r \mu_1\mu_2 x_1 y_1}  F^{6  \ t r  \mu_1 \mu_2 x_1 y_1}  \\
&& + \sqrt{2} \mu_4 {\sqrt{3}\over 2} \gamma_1 \int C_{5 \  t \mu_1\mu_2 x_1 y_1} dt  d\mu_1 d\mu_2. 
\end{eqnarray}
The coefficient of the first term on the rhs above arises as follows. The factor of ${1\over 9}$ arises because the $T^6/(Z_3)^2$ orbifold has  a volume
 which is smaller than that  of  $T^6$ by a factor of $1/9$.
In addition after including the $Z_2$ orientifolding the volume has a factor of $1/2$, this cancells against the factor of $\kappa_{10}^2$in our conventions since $2 \kappa_{10}^2=1$. 


From this action, we get  the equation of motion,
\begin{equation}
 {\gamma_2 \gamma_3 \over 12 } \partial_r \left( \sqrt{-g_4} F_6^{\:0 1 2 r x_1 y_1} \right) = \sqrt{2} \mu_4 \delta(r - r_0).
\end{equation}
Integrating with respect to $r$ then gives the jump in $F_6$
\begin{equation}\label{f6quant}
 \Delta  {F_{6}}_{\:0 1 2 r x_1 y_1} = {r^2 \over R^2} {12 \sqrt{2} \gamma_1 \over \gamma_2 \gamma_3} \mu_4 .
\end{equation}

In our normalization the related component of the Four form flux is given by, 
\begin{equation}
 F_4 = 4 \left(3 \over \kappa \right)^{1 \over 3}  e_1 dx_2 \wedge dy_2 \wedge  dx_3 \wedge dy_3 .
\end{equation}
Using the duality,
\begin{eqnarray}
\nonumber
 {F_{6}}_{\:0 1 2 r x_1 y_1}  &=& \sqrt{g} \: \epsilon_{0 1 2 r x_1 y_1 x_2 y_2 x_3 y_3 } \: F_4^{x_2 y_2 x_3 y_3} \\
&=& \left( r \over R \right)^2 \left(3 \over \kappa \right)^{1 \over 3}  \left( 4  e_1 \gamma_1 \over \gamma_2 .\gamma_3\right),
\end{eqnarray}
then gives,  
\begin{equation}
 \Delta {F_{6}}_{\:0 1 2 r x_1 y_1} = {r^2 \over R^2} \left(  4 \gamma_1 \over \gamma_2 \gamma_3 \right) \left(3 \over \kappa \right)^{1 \over 3}    \delta e_1 .
\end{equation}
Using the jump in $F_6$ calculated in  eq(\ref{f6quant}) then gives the jump in $F_4$ to be,
\begin{equation}
 \label{normjan}
\delta e_1 = 3^{2 \over 3} \sqrt{2} \kappa^{1 \over 3} \mu_4 . 
\end{equation}
For a $D4$ brane which wraps the first $T^2$  $\delta n_1$ times  the rhs would be multiplied by the factor $\delta n_1$. 

More generally consider a $D4$ brane which wraps a two-cycle with wrapping numbers $\delta n_a$. This results in a jump of
the fluxes $\delta e_a$.  
Now  $\delta e_a$ and the winding numbers $\delta n_a$ will  be related by the factor on the rhs of 
eq.(\ref{normjan}) for each $a$.  

The discussion above should actually have been carried out more correctly  by  including various Cherns Simons terms, and working with the 
gauge invariant four-form
$\tilde{F}_4$ and its dual $\tilde{F}_6$.  It is easy to see that doing this gives the same results. 

\subsection{Normalization of the Kahler Form}
Consider the two-cycle of the previous subsection which extends 
 along the first $T^2$
and is located at a point in the other two $T^2$'s. The  $Z_3\times Z_3$ orbifold symmetry  does not result in any self-identifications
on the two-cycle, which has nine images under this symmetry. 

Now in  our notation the  Kahler form is  defined  so that its pull back onto the two-cycle after integration gives, 
\beq
\label{pbint}
\int P[J] = v_1.
\eeq
The actual size of the $T^2$ from the metric eq.(\ref{metric}) is 
\beq
\label{acsizejan}
\int \gamma_1 dx_1 dy_1= {\sqrt{3}\over 2} \gamma_1.
\eeq
Using the relation eq.(\ref{kmoduli}) one gets, 
\beq
\label{actjan}
\int \gamma_1 dx_1 dy_1=3^{2/3} \kappa^{1/3} \int P[J].
\eeq
This shows that with our definition the  integral of the Kahler form must be multiplied by the factor 
$3^{3/2} \kappa^{1/3}$ to get the correct  area of the two-cycle. In effect this means that the relation between the Pull-back of the metric and 
the Kahler form has this extra factor on the lhs of eq.(\ref{ineq}). 
The right hand side of eq.(\ref{lbjan}) must  therefore be multiplied by this extra  factor also to get the correct 
lower bound $T_L$.

\section{Appendix D}

\subsection{ Approximate Symmetry between Type 2) vacua decays and susy vacua decays}
Consider a general Type 2) vacua with all fluxes activated. This vacuum meets the conditions, 
$\text{sign}(m_0\hat{e}_a)>0$ for all $a$. 
Now consider a partner susy vacuum with $\hat{e_a}$ reversed in sign and the same values for all the other fluxes, 
$m_0, p $ and $m_a$.  The ground state energy only depends on $|\hat{e}_a|$. So the  partner vacuum has the same vacuum energy density. 

Now consider a  Type 2) vacuum  decay  mediated by  a brane with charges, $(\delta e_0, \delta e_a, \delta m_a)$.
The resulting  final  vacuum has charges, $e'_a=e_a+\delta e_a$ etc. 
The corresponding values of 
\beq
\label{valijan}
\hat{e}'_{a}=e'_{a}+{\kappa_{abc} \over 2 m_0}(m_b+\delta m_b) (m_c+\delta m_c).
\eeq

Let us relate this to  decay of the partner susy vacuum mediated by the corresponding anti-brane
which  carries opposite charges, 
$(-\delta e_0, -\delta e_a, -\delta m_a)$. 
The  domain walls in the two decays  have the same tension. 
 
A little algebra shows that values of the flux $\hat{e}_a$ in the resulting final vacuum which we denote by 
$\tilde{\hat e}_a'$ is given by, 
\beq
\label{valfjan}
\tilde{\hat e}_a'=-\hat{e}_a' + O(\delta m_a)^2.
\eeq
Thus the two  values in eq.(\ref{valijan}), eq.(\ref{valfjan}) are the same upto corrections of $O(\delta m_a)^2$. 
This is significant because as was mentioned above the ground state energy only depends on $|\hat{e_a}|$. Since this is the same for the two final vacua at  leading order, and since the Type 2) vacuum and its susy partner also have the same ground state energy we find that the energy difference
$\epsilon$ is the same for the two decays.

So to summarize,  we see that for   a Type 2) vacuum   there is a partner susy vacuum. Any decay of the Type 2) vacuum maps to a corresponding decay 
of the susy vacuum, where the tension of the brane mediating the two decays is the same and the energy difference between the true and false 
vacuum for the two decays is also the same to leading order. 
 It then follows that the first decay can occur iff the second one does. Since the susy vacuum is stable one finds that the Type 2) vacuum 
will also not decay, more correctly it can at best  decay  marginally.

\subsection{D$8$-brane and NS$5$-brane mediated decays}
Finally we consider the possibility of the domain wall also containing a
D8 and NS$5$ brane component. We will find that such decays do not  lie within the thin wall approximation.
The $D8$ brane wraps all six directions of the internal space and causes a jump in $F_0$, i.e. in $m_0$.
The $NS5$ brane causes a jump in $H_3$, i.e. in $p$.
Both must change in conjunction so that the tadpole condition eq.(\ref{tadpole}) continues to hold.

A $D8$ brane has tension,
\begin{equation}
 T_8 \sim  \; vol \; e^{-\phi} ({e^{2 \phi} \over vol })^{3 \over 2} \sim |\hat{e}_1 \hat{e}_2 \hat{e}_3|^{-{3 \over 4}}.
\end{equation}
A $NS5$ brane has tension which is comparable,
\begin{equation}
 T_{NS5} \sim  e^{-2\phi} \sqrt{vol} ({e^{2 \phi} \over vol})^{3 \over 2} \sim |\hat{e}_1 \hat{e}_2 \hat{e}_3|^{-{3 \over 4}}.
\end{equation}

The fractional change in $m_0$ that results is of order unity, i.e., ${\delta m_0 \over m_0}\sim 1$. Similarly in $p$.
This causes a  fractional jump in the $v_i$ moduli ${\delta v_i \over v_i}$, also of order unity giving,  for the moduli contribution to the domain wall for the $v_i$ moduli,
\begin{equation}\label{Tmod8}
 T_{v_i\text{mod}} \sim ({\delta v_i \over v_i})^2     {M} \sim |\hat{e}_1 \hat{e}_2 \hat{e}_3|^{-{3 \over 4}}.
\end{equation}
 We see that this is comparable to the $D8$ and $NS5$ contributions.
Thus the thin wall approximation is not good.

\end{document}